\documentclass[%
superscriptaddress,
 amsmath,amssymb,
 aps, physrev,
]{revtex4-2}

\usepackage{hyperref}
\usepackage{graphicx}
\usepackage{dcolumn}
\usepackage{bm}
\usepackage[english]{babel} 
\usepackage[mathlines]{lineno}
\usepackage{verbatim} 

\usepackage[dvipsnames]{xcolor}
\usepackage{cancel}
\usepackage[normalem]{ulem}
\usepackage{soul} 
\sethlcolor{yellow}  

\begin{document}


\title{
Pressure drop-flow rate nonlinearity \\ in bubble trains through a capillary bundle
}

\author{Paolo Botticini}
\email{Contact author: p.botticini@studenti.unibs.it}
\affiliation{PoreLab, Department of Physics, Norwegian University of Science and Technology NTNU, N-7491 Trondheim, Norway\qquad\qquad\phantom{-}}
\affiliation{Department of Mechanical and Industrial Engineering, Università degli Studi di Brescia, Brescia 25123, Italy}

\author{Davide Picchi}
\affiliation{Department of Mechanical and Industrial Engineering, Università degli Studi di Brescia, Brescia 25123, Italy}

\author{Santanu Sinha}
\affiliation{PoreLab, Department of Physics, Norwegian University of Science and Technology NTNU, N-7491 Trondheim, Norway\qquad\qquad\phantom{-}}

\author{Alex Hansen}
\affiliation{PoreLab, Department of Physics, Norwegian University of Science and Technology NTNU, N-7491 Trondheim, Norway\qquad\qquad\phantom{-}}


\begin{center}
\emph{Phys.\ Rev.\ Fluids} \textbf{11}, 073601 -- \href{https://journals.aps.org/prfluids/abstract/10.1103/7z3l-zpzy}{\textbf{Published 6 July, 2026}}
\end{center} 
\vspace{0.2cm}

\begin{abstract}
We investigate the effective rheology of a train of elongated bubbles of negligible viscosity flowing in capillary tubes. Building upon the classical Bretherton theory for a single bubble, we extend the analysis to a train of bubbles in a single capillary tube and finally to an array of parallel, noninteracting capillary tubes, i.e., a capillary bundle. Our goal is to characterize the nonlinear pressure drop-flow rate relation of this simplified two-phase system by incorporating the thin-film hydrodynamics at small capillary numbers. We model the structural heterogeneity of the bundle by assuming that the tube radii follow a truncated power-law distribution and examine deviations of the system from the Darcy law in terms of both its statistical properties and the parameters characterizing the bubble-train (i.e., the tube slenderness ratio, the volume fraction and the number of bubbles). 
The main result is that two-phase flow alters the effective rheology, leading to deviations from Darcy-type behavior across the entire parameter space investigated. Specifically, for a limited number of bubbles, the flow exhibits a smooth transition from the Bretherton regime, where the pressure drop scales with the flow rate to the power of $2/3$, to weaker sublinear regimes with exponents between $2/3$ and unity. Interestingly, increasing the number of bubbles or narrowing the pore-size distribution leads to only minor deviations from the Bretherton regime.  
The resulting pressure drop-flow rate exponents are qualitatively similar to those reported in the literature for immiscible two-phase flow in porous media, despite the inherent simplicity of the capillary bundle model.

\end{abstract}


\maketitle


\section{\label{sec:intro}Introduction}
Two-phase transport processes in porous materials are central to numerous areas of science~\citep{Bear1988, Dullien1992}, including hydrogeology and oil recovery, reservoir and biomechanical engineering, and design of fuel cells and other industrial devices~\citep{JoekarNiasar2010}.
For example, suspensions of air bubbles in water have been used to enhance bioremediation by introducing oxygen into groundwater~\citep{Fry1997}, as well as to remove volatile organic compounds from contaminated soils~\citep{Boehler1990}.
In these systems, the dispersed gas phase alters the flow within individual pores, thereby influencing macroscopic transport properties such as hydraulic conductivity.

In capillaries, gas–liquid mixtures preferentially adopt a plug-train configuration, referred to as Taylor (or segmented, intermittent) flow, where elongated bullet-shaped bubbles are separated by liquid slugs and surrounded by a thin lubricating film~\citep{Suo1964, Woerner2020}.
The incomplete displacement of the wetting phase by the nonwetting discontinuous phase~\citep{PayatakesDias1984}
complicates the pressure drop analysis, making standard two-phase approaches -- such as homogeneous flow models~\citep{Beattie1982, Triplett1999, Kawahara2002} and separated-flow models, e.g., the Lockhart-Martinelli and Chisholm correlations~\citep{Chisholm1967, TaitelDukler1976} -- inadequate. This motivates the use of models that explicitly account for the thin liquid film around the bubble, such as that derived by Bretherton~\citep{Bretherton1961}.
Bretherton~\citep{Bretherton1961} studied the displacement of a viscous fluid by a long bubble of negligible viscosity through a capillary tube of radius $r$ in the lubrication limit. When surface tension dominates viscous stresses (i.e., at low capillary numbers), and in the absence of inertial and gravitational effects, the thickness of the film formed by the advancing bubble -- considered in a reference frame attached to the bubble -- admits a self-similar description. Specifically, the bubble is assumed to be sufficiently long that a central region of uniform film thickness, $h_{\infty} \ll r$, is established. In the front and rear, the shape of the dynamic meniscus is determined by the interplay of viscous and capillary forces, that connects the uniform film to a hemispherical cap (static meniscus), where capillary forces dominate and the curvature remains constant (see Fig.~\ref{fig:pressure_decomp}). This analysis yields the asymptotic scaling laws for the film thickness, the speed ratio between the bubble and the flow ahead, and the overall pressure drop across the bubble.

Later,~\citeauthor{Bretherton1961}'s framework~\cite{Bretherton1961} was extended to describe multiple bubbles (i.e., bubble trains) in smooth capillaries and bead packs~\citep{HirasakiLawson1985, Ratulowski1989}. These models assume joint motion of the gas, the separating liquid slugs, and the wetting films at a common velocity, enabling a pseudohomogeneous single-phase representation of the mixture -- as a chain of bubbles separated by liquid lamellae -- treated as a Darcy flow driven by an applied pressure gradient~\citep{Kornev1999}.
Despite its foundational value,~\citeauthor{Bretherton1961}'s work~\cite{Bretherton1961} is rarely mentioned in studies of gas–liquid pressure drop in capillaries (see, for example,~Sec.\,7 in~\citet{Etminan2021} for an exhaustive review). An exception is the semi-empirical model by~\citet{Kreutzer2005}, which decomposes the total pressure drop in Taylor flow into two main contributions: (i) viscous losses in the liquid slugs, and (ii) the capillary pressure jumps near the bubble caps, based on~\citeauthor{Bretherton1961}'s result for an isolated gas bubble~\cite{Bretherton1961}. 
\citet{Warnier2010} extended the applicability of~\citeauthor{Kreutzer2005}'s model~\cite{Kreutzer2005} to higher capillary numbers by using the scaling arguments of~\citet{Aussillous2000}, and accounted for a more detailed description of the bubble volume. 
However, their analysis remains dimensional, which hampers the identification of two-phase pressure drop-flow rate scaling laws. 
Such a limitation becomes particularly relevant in the context of porous media.

In fact, the motion of bubbles in straight tubes is frequently used as a proxy for understanding two-phase flow at the pore scale~\citep{Olbricht1996}. Although this idealized geometry is an oversimplification of real porous structures, it still offers valuable insights into the hydrodynamic mechanisms at play~\citep{Hunt2003}. 
For instance,~\citet{StarkManga2000} numerically investigated the motion of discrete nonwetting bubbles through a network of parallel straight tubes under an imposed flow rate. They modified the electrical analogy of the network~\citep{Koplik1985} by incorporating \citeauthor{Bretherton1961}'s theory~\cite{Bretherton1961} to evaluate the mean residence time of uniformly distributed bubbles and the effective permeability of the medium. 
By analogy with discrete bubble networks, flow connectivity in porous domains is governed by the evolving phase topology: As the capillary number increases, pathways transition from quasi-static disconnected clusters, to ganglion mobilization, and ultimately to viscous-dominated connected flow~\citep{Schluter2016, Armstrong2016}. These configurations can be mapped onto idealized patterns, from capillary bundles at low capillary numbers to core-annular flow at high capillary numbers~\citep{Picchi2018WRR, Picchi2019, Picchi2018CAFPlug, Picchi2020Geo}.

At the pore scale, steady-state two-phase flow of immiscible Newtonian fluids across a sample of length $L$ can be described by a power-law relation between the effective pressure drop -- defined as the difference between the applied pressure drop $\Delta p$ and a possible threshold pressure $\Delta p_t$ (e.g., a depinning pressure marking the onset of flow~\citep{Xu2014}) -- and the resulting volumetric flow rate $Q$~\citep{Roy2024},
i.e., 
\begin{equation}\label{eq:NonDarcyNew}
    Q^{\zeta}=K_{\text{eff}}\,\dfrac{\Delta{p}-\Delta{p}_{t}}{L},
\end{equation} 
where $K_{\text{eff}}$ is an effective coefficient depending on the flow configuration, and $\zeta\in\left(0,\,1\right)$ is an exponent reflecting the gradual mobilization of fluid-fluid interfaces (note that this form is equivalent to $Q\propto(\Delta p-\Delta p_t)^{1/\zeta}$). In fact, the value of $\zeta$ quantifies the transition from linear Darcy behavior (i.e., $\zeta = 1$)~\citep{Darcy1856} to a sublinear regime caused by the evolving flow pattern~\citep{GaoBlunt2020}. 
Unfortunately, despite extensive studies, see Table~\ref{tab:literature}, a clear quantification of $\zeta$ and its dependence on flow conditions remains an open issue in the community.

\begin{table}[ht]
\centering
\resizebox{\textwidth}{!}{%
\begin{tabular}{|c|c|c|c|c|}
\hline
\textbf{Authors} & \textbf{Method} & \textbf{Wetting and nonwetting phase} & \textbf{Threshold} & \textbf{Power-law exponent} \\
\hline
\citet{Tallakstad2009PRL, Tallakstad2009PRE} & E & glycerin/water and air & no & $0.54\pm0.08$ \\
\hline
\citet{Rassi2011, Rassi2014} & E & water and air & no & $0.3$\,--\,$0.45$ \\
\hline
\citet{Sinha2012} & T, N &  & yes & $0.5$ \\
\hline
\citet{Yiotis2013} & N ($\dagger$) & water and NAPLs & yes & $0.5$ \\
\hline
\citet{Aursjo2014} & E & glycerin/water and rapeseed oil & no & $0.67\pm0.05$, $0.74\pm0.05$ \\
\hline
\citet{Chevalier2015} & E & water and n-heptane & no & $0.65\pm0.1$ \\
\hline
\citet{Sinha2017} & E, N & water and air & yes & $0.46\pm0.05$ -- $0.54\pm0.03$ \\
\hline
\citet{Roy2019} & T, N &  & yes & $1/2$ or $2/3$ \\
\hline
\citet{Yiotis2019} & N ($\dagger$) & water and oil & yes & $2/3$ \\
\hline
\citet{GaoBlunt2020} & E & water and oil & no & $0.60\pm0.01$ \\
\hline
\citet{Fyhn2021} & T, N &  & yes & $1/2$ or $2/3$ \\
\hline
\citet{Roy2021} & T, N &  & yes & $1/2$ or $2/3$ \\
\hline
\citet{Zhang2021} & E & KI brine and n-decane & yes & $0.44\pm0.02$ -- $0.74\pm0.02$ \\
\hline
\citet{Zhang2022} & E & KI brine and n-decane & yes & $0.50\pm0.01$ -- $0.58\pm0.01$ \\
\hline
\citet{Fyhn2023} & N &  & both & $0.39$\,--\,$0.45$ \\
\hline
\citet{Anastasiou2024} & E & water and n-heptane & no & $0.714$\,--\,$1$ \\
\hline
\end{tabular}
}
\caption{Summary of theoretical (T), numerical (N), and experimental (E) studies on the effective rheology of two-phase flows in porous media.
For each study, we report whether a threshold pressure $\Delta p_t$ is included in the model (yes/no/both), together with the estimated exponent $\zeta$ of the power-law pressure drop-flow rate relation, see Eq.~\eqref{eq:NonDarcyNew}.
The symbol ($\dagger$) denotes studies in which the flow is driven by a constant body force, with the Bond number (representing the ratio of gravitational to capillary forces) used as the control parameter instead of the capillary number.
}
\label{tab:literature}
\end{table}

Motivated by this, we investigate the steady-state effective rheology of a two-phase gas–liquid Taylor flow consisting of immiscible Newtonian fluids in a smooth capillary tube and then extend the analysis to a bundle of parallel noninteracting capillary tubes with nonuniform radii reflecting statistical heterogeneity.
Our model shows that, when accounting for the fluid dynamics of a bubble train, this simplified conceptual framework for flow through a porous medium~\citep{Roy2021} exhibits a nonlinear pressure drop-flow rate behavior qualitatively similar to that observed in porous media.

Accordingly, we first present the theoretical framework for decomposing the pressure drop in a liquid-filled tube containing a single elongated gas bubble, reviewing existing correlations~(Sec.\,\ref{sec:single}). We then extend this analysis to arbitrary bubble trains and recast the problem in dimensionless form~(Sec.\,\ref{sec:tube}), identifying the key parameters governing the total pressure drop across the tube~(Sec.\,\ref{sec:parameter}).
Next, we examine how the pressure drop varies across the parameter space~(Sec.\,\ref{sec:pressure-flow}), and identify the model's applicability region~(Sec.\,\ref{sec:applicability}). In the slender-tube limit, perturbation analysis reveals how a single gas bubble alters the system dynamics~(Sec.\,\ref{sec:asymptotics}).
Finally, we extend this framework to a capillary bundle by introducing a statistical distribution of pore sizes. After recasting the problem in dimensionless form~(Sec.\,\ref{sec:bundle_adim}), we show that our simplified model captures the same qualitative nonlinear pressure drop–flow rate scaling typical of two-phase flow in porous media~(Sec.\,\ref{sec:2phasebundle}), despite neglecting key complexities of real pore networks, such as pore connectivity, contact-line friction, and pinning dynamics.
A summary discussion is provided in the concluding section~(Sec.\,\ref{sec:conclusion}).

\section{Single-capillary model}\label{sec:STM}
\subsection{\label{sec:single}Pressure drop decomposition across an elongated bubble} 
We first examine the case of a single elongated bubble moving in a {round} capillary tube, focusing on the visco-capillary regime where inertia and buoyancy effects are negligible. In~Sec.\,\ref{sec:tube}, this theoretical framework is extended to account for multiple bubbles within the capillary.

\begin{figure}
    \centering
    \includegraphics[width=0.75\linewidth]{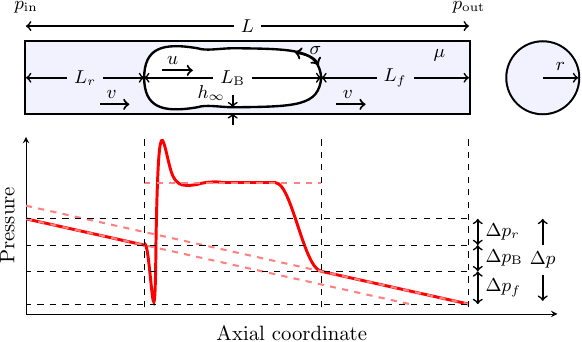}
    \caption{
    {An elongated bubble of length $L_{\rm{B}}$ travels at constant speed $u$ through a capillary tube of radius $r$ and length $L$, displacing a liquid of viscosity $\mu$, and average speed $v$, with interfacial tension $\sigma$. In its central portion, the film thickness is uniform and equal to $h_{\infty}$. The flow is driven by a pressure difference $\Delta{p}=p_{\rm{out}}-p_{\rm{in}}$, which can be decomposed as shown in the plot of the axial pressure distribution, see Eq.~\eqref{dptot_dim}: the red dashed lines indicate the constant pressure in the uniform‑film region and the linear pressure variations within the liquid slugs of lengths $L_{f}$ and $L_{r}$, located ahead of and behind the bubble, respectively.}
    }
    \label{fig:pressure_decomp}
\end{figure}

The bubble, of length $L_{B}$ and negligible viscosity, displaces a liquid with dynamic viscosity $\mu$ along a horizontal smooth capillary tube of radius $r$ and length $L$, as shown in Fig.~\ref{fig:pressure_decomp}.
The flow is driven by an imposed pressure drop $\Delta{p}=p_{\rm{in}}-p_{\rm{out}}>0$ between the inlet and the outlet of the tube, and we assume full wetting of the tube walls by the continuous phase.
The axial pressure drop over the tube length $L$ is decomposed as follows~\citep{Chung2004, Kreutzer2005, Walsh2009, Warnier2010, Gupta2010, Baroud2010, Minagawa2013, MacGiollaEain2015, Ni2017, Kawahara2020, Kurimoto2017, Kurimoto2019, Kurimoto2024}:
\begin{linenomath}
    \begin{equation}\label{dptot_dim}
        \Delta{p}=\Delta{p}_{r}+\Delta{p}_{\rm{B}}+\Delta{p}_{f},
    \end{equation}
\end{linenomath}
where the first and third term on the right-hand side indicate the viscous pressure drop of the liquid slugs respectively at the rear ($r$) and at the front ($f$) of the bubble, and the second term is the pressure drop over the bubble (B), as shown in Fig.~\ref{fig:pressure_decomp}.
Equation~\eqref{dptot_dim} should also account for the viscous losses in the thin liquid film that separates the bullet-shaped bubble from the tube wall. However, when gravity and inertial effects are neglected, and the bubble is treated as inviscid, the film is almost stagnant and does not contribute to the pressure drop~\citep{Gupta2010, Warnier2010, Fouilland2010}. Specifically, in the region of uniform film thickness, the pressure is almost constant, whereas large pressure oscillations can form at the rear of the bubble due to typical meniscus oscillations at the gas-liquid interface~\citep{Kreutzer2005, Ni2017}. 
The pressure drop due to viscous losses in the gas bubble is not taken into account, due to the low viscosity of the gas phase compared to that of the liquid~\citep{Kreutzer2005, Warnier2010}. 

The flow established along the slug regions of length $L_{r}$ and $L_{f}$, located respectively at the rear and front of the bubble, is assumed to be laminar and fully developed.
Under these approximations~\citep{Ni2017}, whose validity is discussed in~Sec.\,\ref{sec:applicability}, the Hagen-Poiseuille equation~\citep{Hagen1839, Poiseuille1846} linearly relates the volumetric flow rate of the liquid $Q$ to the local pressure drop due to viscous friction:
\begin{linenomath}
    \begin{equation}\label{Q_dim_sp}
        Q=\dfrac{\pi\,r^{4}}{8\,L_{j}\,\mu}\,\Delta{p}_{j}\,,\qquad{j}=\left\{r,\,f\right\}.
    \end{equation}
\end{linenomath}
Thus, the local pressure drop can be expressed in terms of the average speed of the liquid slugs $v$, i.e., $Q=\pi\,r^{2}\,v$, as
\begin{linenomath}
    \begin{equation}\label{dpj_dim}
        \Delta{p}_{j}=\dfrac{8\,L_{j}\,\mu}{r^2}\,v\,,\qquad{j}=\left\{r,\,f\right\}.
    \end{equation}
\end{linenomath}
The present analysis assumes that the viscous flow in the liquid slugs ahead of and behind the bubble can be treated independently, so their pressure drops are additive. Accordingly, the overall pressure drop across the liquid phase is obtained by summing the two contributions in Eq.~\eqref{dpj_dim}:
\begin{linenomath}
    {
    \begin{equation}\label{eq:new_slug}
        \Delta{p}_{\rm{S}}=\Delta{p}_{r}+\Delta{p}_{f}=\dfrac{8\left(L_{r}+L_{f}\right)\,\mu}{r^2}\,v
        =\dfrac{8\left(L-L_{\rm{B}}\right)\,\mu}{r^2}\,v,
    \end{equation}
    }
\end{linenomath}
where the sum of the rear and front slug lengths has been expressed as the difference between the {tube} and the bubble length, assuming that both phases are incompressible.

The presence of the gas bubble causes an excess pressure drop~\citep{Warnier2010, Etminan2021}, $\Delta{p}_{\rm{B}}$ in Eq.~\eqref{dptot_dim}, expressed in terms of the capillary number based on the bubble speed $u$:
\begin{linenomath}
    \begin{equation}\label{capillary}
        \text{Ca}=\dfrac{\mu\,u}{\sigma},
    \end{equation}
\end{linenomath} 
which can be interpreted as the ratio between viscous stresses and {interfacial} tension, $\sigma$. 
Since we are interested in the visco-capillary regime, we restrict our analysis to $\text{Ca}\ll1$ (in the range $\text{Ca}\le 10^{-2}$) where the tip-to-tip pressure drop is independent of the bubble length and proportional to the capillary pressure $\sigma/r$. In this regime, the pressure drop can be decomposed into two contributions: (i) the dynamic jump due to the curvature of the front and rear menisci, first described by~\citeauthor{Bretherton1961}, which dominates at small capillary numbers and scales as $\text{Ca}^{2/3}$~\cite{Klaseboer2014, Langewisch2015, Cherukumudi2015}; (ii) the jump in normal viscous stresses,
incorporated in the empirical model of~\citet{Balestra2018}, which becomes important for $\text{Ca}\gtrsim 10^{-3}$. This contribution takes the form of a ratio of polynomials in $\text{Ca}$ and higher powers.
Accordingly, we perform an asymptotic expansion in powers of $\left(3\,\text{Ca}\right)^{1/3}$ of~\citeauthor{Balestra2018}'s model retaining terms up to $\mathcal{O}(\text{Ca})$ yielding
\begin{linenomath}
    \begin{subequations}\label{dpb_dim}
    \begin{align}
        \Delta{p}_{\rm{B}}&=\dfrac{\sigma}{r}
        {\big[\underbrace{\beta_{2}\left(3\,\text{Ca}\right)^{2/3}}_{\text{(i)}}
        +\,\underbrace{\beta_{3}\left(3\,\text{Ca}\right)}_{\text{(ii)}}}
        \big],\quad{\text{for}\quad (3\,\text{Ca})^{2/3}\ll1,}\\
        &{\text{with}\qquad\beta_{2}=4.914,\qquad\beta_{3}=-4.391,}
    \end{align}
    \end{subequations}
\end{linenomath}
where $\beta_{2}$ and $\beta_{3}$ are constant fitting parameters determined from the model of~\citet{Balestra2018}.
Note that Eq.~\eqref{dpb_dim} includes terms up to $\mathcal{O}(\mathrm{Ca})$ rather than only Bretherton's contribution [term~(i)], because the pressure drop in the liquid slug in Eq.~\eqref{dptot_dim}, once normalized, scales as the capillary number (see Sec.~\ref{sec:tube}). Therefore, our model for the pressure drop is consistent up to $\mathcal{O}(\mathrm{Ca})$.

\subsection{Extension to bubble train and dimensionless formulation}\label{sec:tube}
\begin{figure}
    \centering
    \includegraphics[width=1\linewidth]{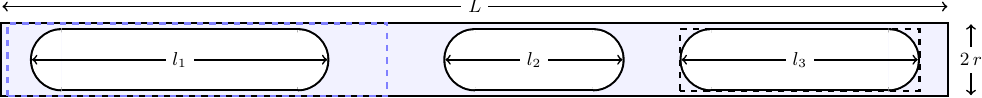}
    \caption{
    Schematic of a bubble train with $N=3$ gas bubbles separated by liquid slugs in a capillary tube. The front and rear menisci of each bubble are modeled as hemispherical caps.
    The blue dashed rectangle represents one element of the repeating bubble–slug pattern.
    }
    \label{fig:train}
\end{figure}

Building on the analysis for a single bubble, we extend the pressure drop model to a train of $N$ bubbles of individual lengths $l_{i}$, see Fig.~\ref{fig:train}. For the liquid phase, we replace $L_{\rm{B}}$ in Eq.~\eqref{eq:new_slug} with the total length occupied by the gas, $\sum_{i\le{N}}l_{i}$, while for the bubble contribution we multiply Eq.{~\eqref{dpb_dim}} by $N$, yielding the overall pressure drop
\begin{linenomath}
    {
    \begin{equation}\label{eq:new_dim_p}
        {\Delta{p}
        =\underbrace{\dfrac{8\left(L-\sum_{i\le{N}}l_{i}\right)\,\mu}{r^2}\,v}_{\Delta{p}_{\rm{S}}}\,
        +
        \,\underbrace{N\,\dfrac{\sigma}{r}
        \big[\beta_{2}\,\left(3\,\text{Ca}\right)^{2/3}
        +\,\beta_{3}\left(3\,\text{Ca}\right)\big]}_{\Delta{p}_{\rm{B}}}.}
    \end{equation}
    }
\end{linenomath}
We normalize Eq.~\eqref{eq:new_dim_p} using the capillary pressure scale, $\mathcal{P}={\sigma}\slash{{r}}$, which includes exclusively parameters related to the system geometry and interfacial properties and it is consistent with the small-capillary-number regime. 
The choice of alternative length scales in the definition of the pressure scale $\mathcal{P}$ is discussed in Sec.~\ref{sec:asymptotics}.
We denote the resulting dimensionless quantities by the asterisk, i.e., $\Delta{p}^{\ast}=\Delta{p}\,\mathcal{P}^{-1}$, and introduce the reduced capillary number 
\begin{linenomath}
    \begin{equation}\label{eq:zCa}
        z=\left(3\,\rm{Ca}\right)^{1\slash3},
    \end{equation}
\end{linenomath}
to lighten notation.
This procedure yields
\begin{linenomath}
    \begin{equation}\label{eq:new_adim_drop}
        {\Delta{p}^{\ast}
        =\underbrace{\dfrac{8}{3}\,z^3\,
            \dfrac{v}{u}\,\dfrac{L}{r}\,\left(1-\varphi\right)}
        _{\Delta{p}_{\rm{S}}^{\ast}}\,
        +
        \,\underbrace{N
        \big(\beta_{2}\,z^2
        +\,\beta_{3}\,z^3\big)}_{\Delta{p}_{\rm{B}}^{\ast}},}
    \end{equation}
\end{linenomath}
where we have introduced the bubble length fraction as
\begin{linenomath}
    \begin{equation}\label{eq:phi_b}
        \varphi=\dfrac{\sum_{i\le{N}}l_{i}}{L}.
    \end{equation}
\end{linenomath}
In Eq.~\eqref{eq:phi_b}, the speed ratio between the liquid slug and the bubble $v/u$ can be expressed as a function of the uniform film thickness, $h_{\infty}$, normalized by the tube radius, by applying a mass balance in the bubble reference frame, see Refs.~\citep{Stone2010, Langewisch2015, Balestra2018, Picchi2021}, yielding
\begin{linenomath}
    \begin{equation}\label{vuw_relation}
        \dfrac{v}{u}={\left(1-\dfrac{h_{\infty}}{r}\right)^2}.
    \end{equation}
\end{linenomath}
Following the analysis of~\citet{Bretherton1961}, the dimensionless film thickness is given by
\begin{linenomath}
\begin{equation}\label{hinfdivr}
        \dfrac{h_{\infty}}{r}={{0.643}\,\left(3\,\text{Ca}\right)^{2/3}},
    \end{equation}
\end{linenomath} 
where additional corrections, as proposed in the models of~\citet{Aussillous2000,Klaseboer2014,Balestra2018} are not considered here, since they become relevant only at finite capillary numbers.
Therefore, in the limit of small capillary numbers, the bubble moves faster than the mean flow of the liquid on either side, i.e., $u>v$ (see Refs.~\citep{Abiev2008, Langewisch2015}), and the speed ratio remains of order one, with additional terms of order $\mathcal{O}(\text{Ca}^{2/3})$, i.e., $v/u \sim 1 - 1.286\,z^2+\ldots$\,.

Finally, we make the tube length in Eq.~\eqref{eq:new_adim_drop} dimensionless by introducing the slenderness ratio
\begin{linenomath}
    \begin{equation}\label{eq:newLast}
        L^{\ast}=\dfrac{L}{r},
    \end{equation}
\end{linenomath}
yielding the pressure drop model consistent up to $\mathcal{O}(\text{Ca})$, i.e., $\mathcal{O}(z^3)$:
\begin{linenomath}
    \begin{equation}\label{dp_adim_totN_bis}
        \Delta{p}^{\ast}={\dfrac{8}{3}\,z^3\,L^{\ast}
        \left(1-\varphi\right)
        +
        N\,
        \big(\beta_{2}\,z^2
        +\,\beta_{3}\,z^3\big)}{,\qquad \text{for}\quad z^2\ll1}.
    \end{equation}
\end{linenomath}

\subsection{Analysis of the parameter space}\label{sec:parameter}
We now delineate the validity of Eq.~\eqref{dp_adim_totN_bis} by identifying its geometric constraints and provide an equivalent formulation of the model that simplifies the interpretation of the results.

\subsubsection{Bubble length and capillary number}\label{subsec:Ca}
\begin{figure}
    \centering
    \includegraphics[width=1\linewidth]{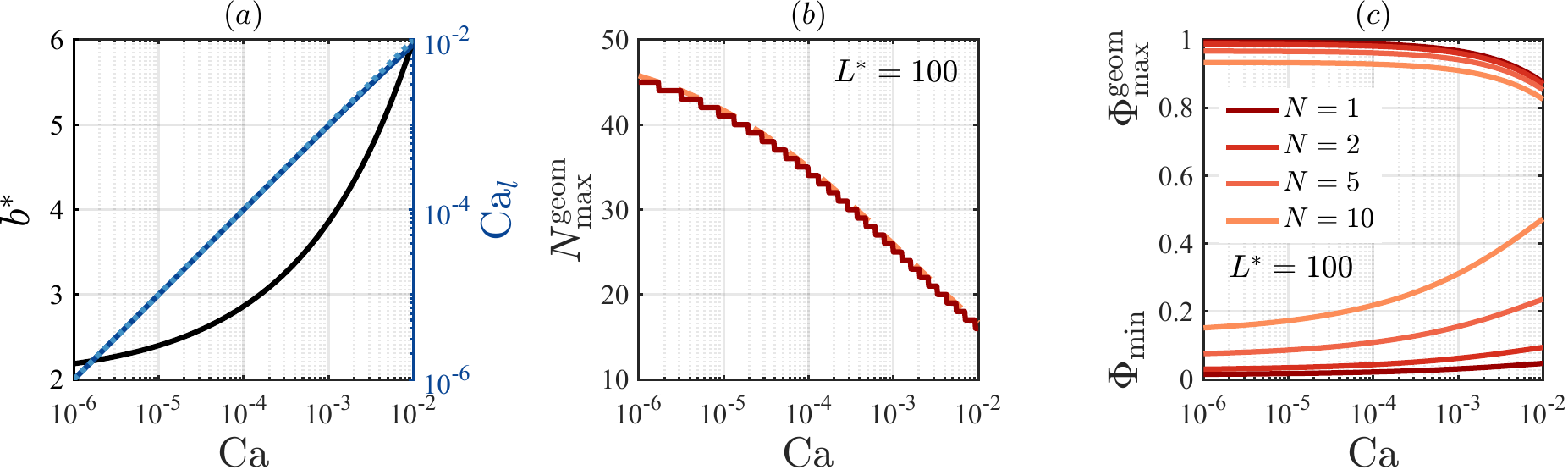}
    \caption{
    ($a$) Dimensionless minimum bubble length $b^{\ast}$ 
    (left ordinate) and capillary number based on the average speed of the liquid slug (right ordinate), as functions of the capillary number based on the bubble speed, see Eq.~\eqref{eq:bastmin} and~\eqref{eq:CalCa}. 
    ($b$) Variation of the maximum bubble number $N_{\rm{max}}^{\rm{geom}}$ with the capillary number, see Eq.~\eqref{eq:Nmax_bubble}, for a slenderness ratio $L^{\ast}=100$, with the superimposed staircase profile resulting from the floor operator.
    ($c$) Geometric range of variation of the gas volume fraction $\Phi$ with the capillary number, see Eqs.~(\ref{eq:Phi_min},\,\ref{eq:Phi_max}), for a slenderness ratio $L^{\ast}=100$ and a varying number $N<N_{\rm{max}}^{\rm{geom}}$ of bubbles in the train.
    }
    \label{fig:CaCal}
\end{figure}

The Bretherton model is applicable only if the bubble is sufficiently long so a region of uniform film thickness exists and the bubble assumes the typical bullet shape. Following~\citet{Klaseboer2014}, this constraint is satisfied when the length of each dynamic meniscus exceeds approximately $\lambda\sim10$ dimensionless units, ensuring that the front and rear spherical caps do not interact and the flow on either side of the bubble can be treated independently. Such condition (see Ref.~\citep{Cherukumudi2015}) can be expressed as  
\begin{linenomath}
	\begin{equation}
		\dfrac{l_{i}}{r}\ge\dfrac{b}{r}{\,=\,}{b}^{\ast},\quad{\text{with}} \quad{b}^{\ast}=2\left[1+\lambda\,\dfrac{h_{\infty}}{r}\left(3\,\text{Ca}\right)^{-1/3}\right]=2\left(1+{0.643}\,\lambda\,{z}\right),
            \label{eq:bastmin}
	\end{equation}
\end{linenomath}
where the last equality follows from Eq.~\eqref{hinfdivr}. As shown in Fig.~\ref{fig:CaCal}\,($a$), Eq.~\eqref{eq:bastmin} highlights that the minimum bubble length scales as $ 2\,r + 1.286\,\lambda\,r\,\text{Ca}^{1/3}$, reducing to the tube diameter in the limit $\text{Ca}\to 0$.

To facilitate comparison with the single-phase limit, we also introduce the capillary number based on the average speed of the liquid, $\text{Ca}_{l}$, and determine its relation with the capillary number based on the bubble speed using Eqs.~\eqref{vuw_relation} and~\eqref{hinfdivr}:
\begin{linenomath}
    \begin{equation}\label{eq:CalCa}
        \text{Ca}_{l}=\frac{\mu\,v}{\sigma}=
        \frac{\mu\,u}{\sigma}
        \frac{v}{u}
        =\text{Ca}\,{\left(1-0.643\,\left(3\,\text{Ca}\right)^{2/3}\right)^{\!2},
        \qquad\text{for}\quad\left(3\,\text{Ca}\right)^{2/3}\ll1,
        }
    \end{equation}
\end{linenomath}
which indicates that the two capillary numbers differ only by terms of $\mathcal{O}(\text{Ca}^{5/3})$ and, therefore, are asymptotically equivalent, as shown in Fig.~\ref{fig:CaCal}\,($a$).

\subsubsection{Volume fraction}
Since mass conservation in multiphase flow is typically expressed in terms of volumes~\citep{Bear1988} rather than lengths, we introduce the gas volume fraction (or saturation)
\begin{linenomath}
    \begin{equation}\label{eq:Phigrande}
        \Phi=\dfrac{\sum_{i\le{N}}V_{i}}{V},
    \end{equation}
\end{linenomath}
where $V_i$ is the volume of the $i$-th bubble and $V$ is the total volume of the tube. This quantity can be related to the linear gas fraction $\varphi$ defined in Eq.~\eqref{eq:phi_b} based solely on geometric considerations. 
Following~\citet{Warnier2010}, estimating each bubble volume as that of a cylinder of length $l_{i}$ and radius equal to that of the bubble in the region of uniform film thickness, i.e., $r_{b}=r-h_{\infty}$, tends to overestimate the actual gas volume. 
In fact, bubbles in confined channels exhibit a bullet-shaped profile, with rounded caps that contain liquid instead of gas. To correct the volume accordingly, the cylinder length is reduced to $l_{i}-\delta$, with $\delta={2}/{3}\,r_{b}$ in the case of hemispherical caps, yielding
\begin{linenomath}
    \begin{equation}\label{eq:phiPhi}
    {
    \Phi=\varphi\left(1-\dfrac{h_{\infty}}{r}\right)^{\!2}-\dfrac{2}{3}\,N\,\dfrac{r}{L}\left(1-\dfrac{h_{\infty}}{r}\right)^{\!3}
    }.
    \end{equation}
\end{linenomath}
By taking the small-$\text{Ca}$ limit and using Eq.~\eqref{eq:newLast}, the previous relation reduces to
\begin{linenomath}
    \begin{equation}\label{eq:phiPhi_new}
        {
        \Phi=\varphi-\dfrac{2}{3}\,\dfrac{N}{L^{\ast}},{\qquad \text{for}\quad z^2\ll1,}
        }
    \end{equation}
\end{linenomath}
where the ratio $N/L^{\ast}$ represents the dimensionless bubble density in the capillary tube, defined as the ratio between the number of bubbles and the tube length normalized using the tube radius. In the limit of a slender capillary tube, $L^{\ast}\to\infty$, the volumetric correction due to the spherical shape of the bubble menisci vanishes, and the bubble length fraction, $\varphi$, and the bubble volume fraction, $\Phi$, become equivalent.

To proper describe the train of elongated bubbles as shown in Fig.~\ref{fig:train}, we impose the following geometric constraints. First, the maximum number of bubbles ($N=N_{\rm{max}}$) is determined looking at a capillary tube filled with tightly packed bubbles, each of them having the minimum admissible length given in Eq.~\eqref{eq:bastmin}. This yields the maximum number of bubbles as
\begin{linenomath}
    \begin{equation}\label{eq:Nmax_bubble}
        \displaystyle
        N_{\rm{max}}^{\rm{geom}}=
        \underset{\rm{Ca}}{\operatorname{min}}
        \left\lfloor\dfrac{L}{b}\right\rfloor=
        \underset{\rm{Ca}}{\operatorname{min}}
        \left\lfloor\dfrac{L^{\ast}}{b^{\ast}}\right\rfloor,
    \end{equation}
\end{linenomath}
where the floor function enforces the number of bubbles to be an integer number, and the minimization over $\text{Ca}$ is needed because the capillary number determines the minimum bubble length for the entire train. For a capillary tube with $L^\star=100$, the evolution of $N_{\rm{max}}^{\rm{geom}}$ with the capillary number is shown in Fig.~\ref{fig:CaCal}\,($b$): As expected, the bubbles lengthen with increasing capillary number, resulting in a decrease in the maximum number of bubbles.

Combining the constraints on the minimum length and the maximum number of bubbles, the resulting saturation is constrained, as shown in Fig.~\ref{fig:CaCal}\,($c$) as a function of the capillary number, for different numbers of bubbles in the train.
The minimum volume fraction corresponds to $N$ bubbles of minimum admissible length and is obtained by substituting $\varphi=N\,b/L$ into Eq.~\eqref{eq:phiPhi}, leading to
\begin{linenomath}
    \begin{equation}\label{eq:Phi_min}
        \Phi_{\rm{min}}=\dfrac{N}{L^{\ast}}
        \left[b^{\ast}-\dfrac{2}{3}\left(1-\dfrac{h_{\infty}}{r}\right)\right]
        \left(1-\dfrac{h_{\infty}}{r}\right)^{2},\qquad\text{with}\quad N\in\left[0;\,N_{\rm{max}}\right].
    \end{equation}
\end{linenomath}
Conversely, the maximum volume fraction is obtained when $\varphi=1$:
\begin{linenomath}
    \begin{equation}\label{eq:Phi_max}
        \Phi_{\rm{max}}^{\rm{geom}}=
        \left[1-\dfrac{2}{3}\,\dfrac{N}{L^{\ast}}\!\left(1-\dfrac{h_{\infty}}{r}\right)\right]
        \left(1-\dfrac{h_{\infty}}{r}\right)^{2},\qquad\text{with}\quad N\in\left[1;\,N_{\rm{max}}\right].
    \end{equation}
\end{linenomath}
These geometric estimates assume ideal packing and do not account for hydrodynamic interactions in the liquid regions ahead of and behind each bubble, which become important before the bubbles are densely packed, as discussed in~Sec.\,\ref{sec:applicability}.

\section{Results and discussion}\label{sec:R&D}
\subsection{Pressure drop-flow rate curves}\label{sec:pressure-flow}
By replacing the linear fraction $\varphi$ with the gas volume fraction $\Phi$ using Eq.~\eqref{eq:phiPhi_new} into Eq.~\eqref{dp_adim_totN_bis}, we obtain the dimensionless two-phase pressure drop for small capillary numbers:
\begin{linenomath}
    \begin{equation}\label{eq:dpast_new}
        {\Delta{p}^{\ast}
        =\underbrace{\dfrac{8}{3}\,z^3\,
        L^{\ast}\left(1-\Phi-\dfrac{2}{3}\,\dfrac{N}{L^{\ast}}\right)}
        _{\Delta{p}_{\rm{S}}^{\ast}}\,
        +
        \,\underbrace{N
        \big(\beta_{2}\,z^2
        +\,\beta_{3}\,z^3\big)}_{\Delta{p}_{\rm{B}}^{\ast}}}
        {,\qquad \text{for}\quad z^2\ll1,}
    \end{equation}
\end{linenomath}
\begin{figure}
    \centering
    \includegraphics[width=1\linewidth]{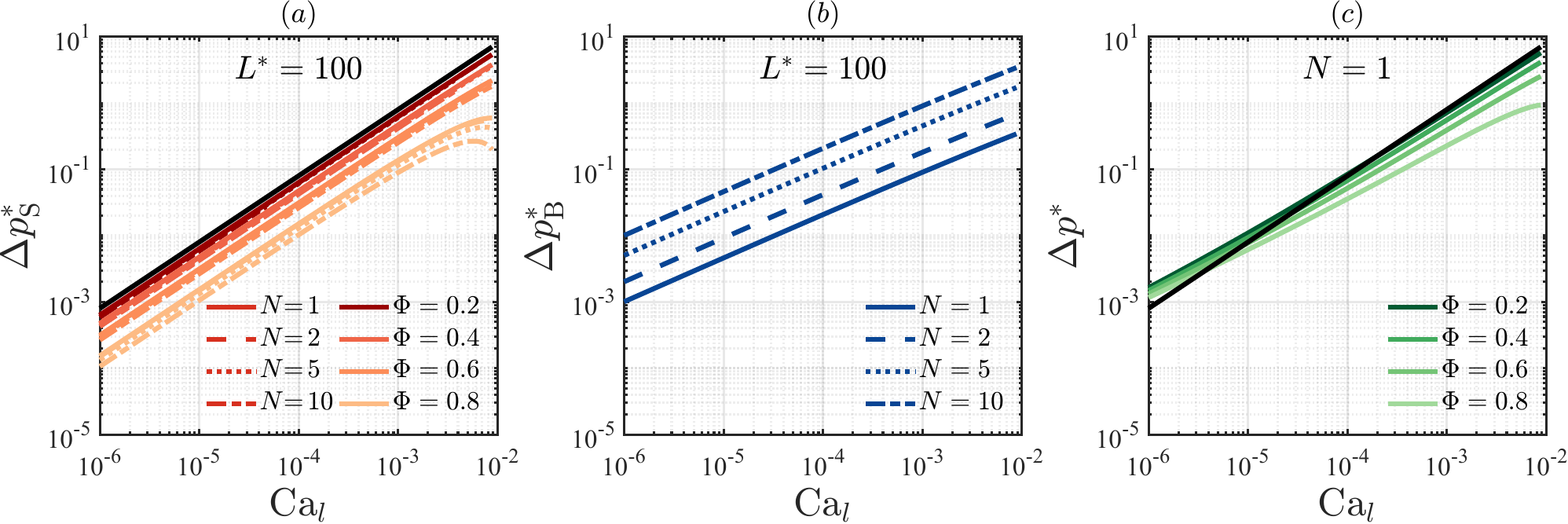}\\[1em]
    \includegraphics[width=1\linewidth]{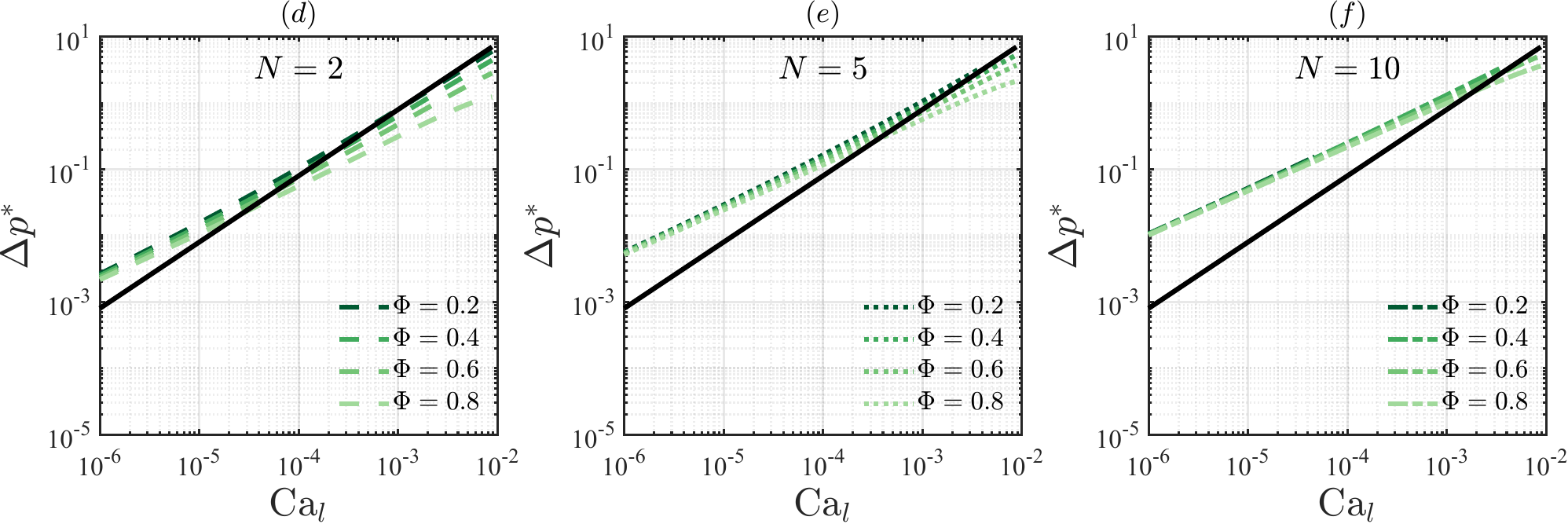}
   \caption{
        Dimensionless pressure drops as functions of the capillary number based on the average speed of the liquid for a slenderness ratio of $L^{\ast}=100$. ($a$), ($b$) Slug- and bubble-related contributions, respectively, varying the gas volume fraction $\Phi$ and the bubble number $N$. Panels ($c$)--($f$) Total pressure drop for fixed bubble numbers and varying the gas volume fraction; for reference, the single-phase limit is shown as a solid black line.
        }
    \label{fig:pressure}
\end{figure}
where the the slug- and bubble-related contributions have been identified. In Fig.~\ref{fig:pressure} we show the evolution of the pressure drop across the slug and the bubbles -- Figs.~\ref{fig:pressure}\,($a$) and \ref{fig:pressure}\,($b$) -- and the total pressure drop -- Figs.~\ref{fig:pressure}\,($c$)--\ref{fig:pressure}\,($f$) -- as functions of the capillary number of the liquid.
The pressure drop in the liquid slugs depends on both the bubble number and the saturation, whereas the bubble-related contribution depends on the bubble number only. 
In Fig.~\ref{fig:pressure}\,($a$), four families of curves are shown for increasing saturation, each identified by a different color, while the linestyle indicates the bubble number, ensuring that the geometric constraint given in Eq.~\eqref{eq:Nmax_bubble} is satisfied and reflecting that the same saturation can be achieved with a larger number of shorter bubbles.
To facilitate comparison with the single-phase limit (i.e., $N,\,\Phi\to0$), shown as a solid black line and consistent with Darcy law, we present these results in terms of the capillary number based on the speed of the liquid using Eq.~\eqref{eq:CalCa}. 
The slug-related pressure drop shown in Fig.~\ref{fig:pressure}\,($a$) decreases as both the bubble volume fraction and the bubble number increase, with small deviations from linearity arising only at larger capillary numbers and volume fractions.
The bubble-related pressure drop shown in Fig.~\ref{fig:pressure}\,($b$) follows the $2/3$ scaling typical of the~\citeauthor{Bretherton1961} problem, with the bubble number $N$ entering the pressure drop only as a multiplicative factor.

Figures~\ref{fig:pressure}\,($c$)--($f$) show the overall pressure drop for a fixed number of bubbles at varying volume fractions. This means looking at a change in bubble length while keeping the number of bubbles and the bubble-induced pressure drop fixed. 
Each bubble increases the gas-related pressure drop, but by shortening the liquid slug it reduces the liquid-phase contribution.
The gas-related pressure contribution is proportional to the number of bubbles, but its growth with $\text{Ca}\sim\text{Ca}_{l}$ follows a $2/3$ power law, slower than the liquid phase contribution, which grows linearly. This marks a transition in the dominant pressure drop, corresponding to the point where the two-phase pressure drop curve intersects the single-phase limit.
Increasing the number of bubbles shifts this transition to higher capillary numbers and reduces its dependence on saturation.

\subsection{Applicability region of the model}\label{sec:applicability}
\begin{figure}
    \centering
    \includegraphics[width=1\linewidth]{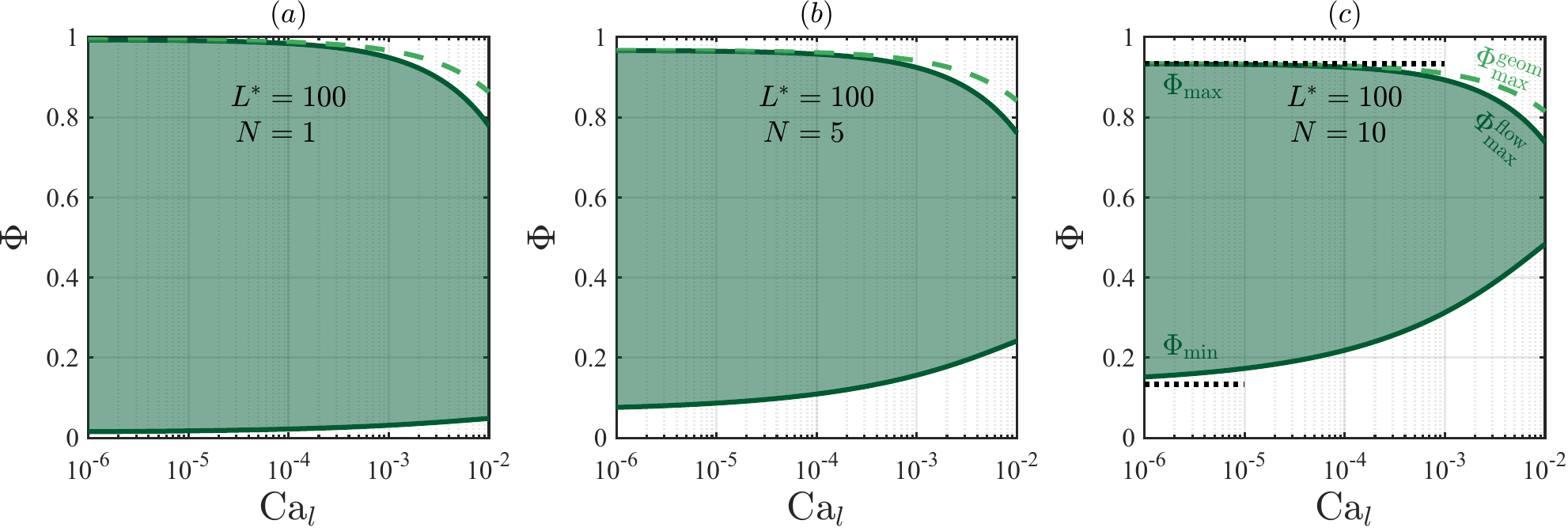}
    \caption{
    Applicability map of the pressure drop model in a capillary tube of slenderness ratio $L^{\ast}=100$, for different numbers of bubbles in the train: ($a$) $N=1$, ($b$) $N=5$, and ($c$) $N=10$.  
    The region shaded in green indicates the parameter space in the plane of capillary number (based on the average speed of the liquid) and gas volume fraction $\left(\text{Ca}_{l}{,}\,\Phi\right)$ where the model is valid. 
    The maximum gas saturation $\Phi_{\rm{max}}^{\rm{flow}}$ due to the monotonicity constraint, see Eq.~\eqref{eq:monotonicity}, and the minimum gas saturation $\Phi_{\rm{min}}$, see Eq.~\eqref{eq:Phi_min}, are shown as solid dark green lines; the upper geometric bound $\Phi_{\rm{max}}^{\rm{geom}}$, see Eq.~\eqref{eq:Phi_max}, as a dashed light green line. 
    In ($c$), the asymptotes corresponding to Eq.~\eqref{eq:limit} are shown as black dotted lines.
    }
    \label{fig:Validity_abc}
\end{figure}

To ensure physical consistency, we require that the dimensionless pressure drop within the slug, $\Delta{p^{\ast}_{\rm{S}}}$, remains an increasing function of the capillary number, imposing the constraint
\begin{linenomath}
    \begin{equation}\label{eq:monotonicity}
         \dfrac{\partial\,{\Delta{p^{\ast}_{\rm{S}}}}}{\partial{z}}>0.
    \end{equation}
\end{linenomath}
This condition reflects the assumption of hydrodynamic independence between adjacent liquid slugs~\citep{Vanapalli2009, Baroud2010}: when the derivative becomes negative, the pressure drop-flow rate curve loses its monotonicity, signaling the onset of bubble interaction.
Therefore, developing this condition from Eq.~\eqref{eq:new_adim_drop} provides an upper bound $\Phi_{\textnormal{max}}^{\textnormal{flow}}$ on the bubble volume fraction more restrictive than the purely geometric limit expressed by Eq.~\eqref{eq:Phi_max}, reducing the maximum number of bubbles that can fit within the capillary, $N_{\rm{max}}<N_{\rm{max}}^{\rm{geom}}$. 
Figure~\ref{fig:Validity_abc} shows how the model's applicability region, defined by the admissible gas volume fractions, varies with the capillary number based on the average speed of the liquid slug, for different numbers of bubbles in the train.   
For small capillary numbers, the geometric and hydrodynamic upper limits converge, while the lower limit is obtained from Eq.~\eqref{eq:Phi_min} assuming bubbles of minimum length:
\begin{linenomath}
    \begin{equation}\label{eq:limit}
        {\Phi_{\textnormal{max}}\to\,}1-\dfrac{2}{3}\,\dfrac{N}{L^{\ast}},
        \qquad
        {\Phi_{\textnormal{min}}\to\,}\dfrac{4}{3}\,\dfrac{N}{L^{\ast}}
        {,\qquad \text{for}\quad z^2\ll1}
        .
    \end{equation}
\end{linenomath}
In the limit of an infinitely long capillary tube, $L^{\ast}\to\infty$, the admissible saturation spans the full range $0<\Phi<1$.

\subsection{Perturbation analysis around the single-phase limit}\label{sec:asymptotics}

\begin{figure}
    \centering
    \includegraphics[width=1\linewidth]{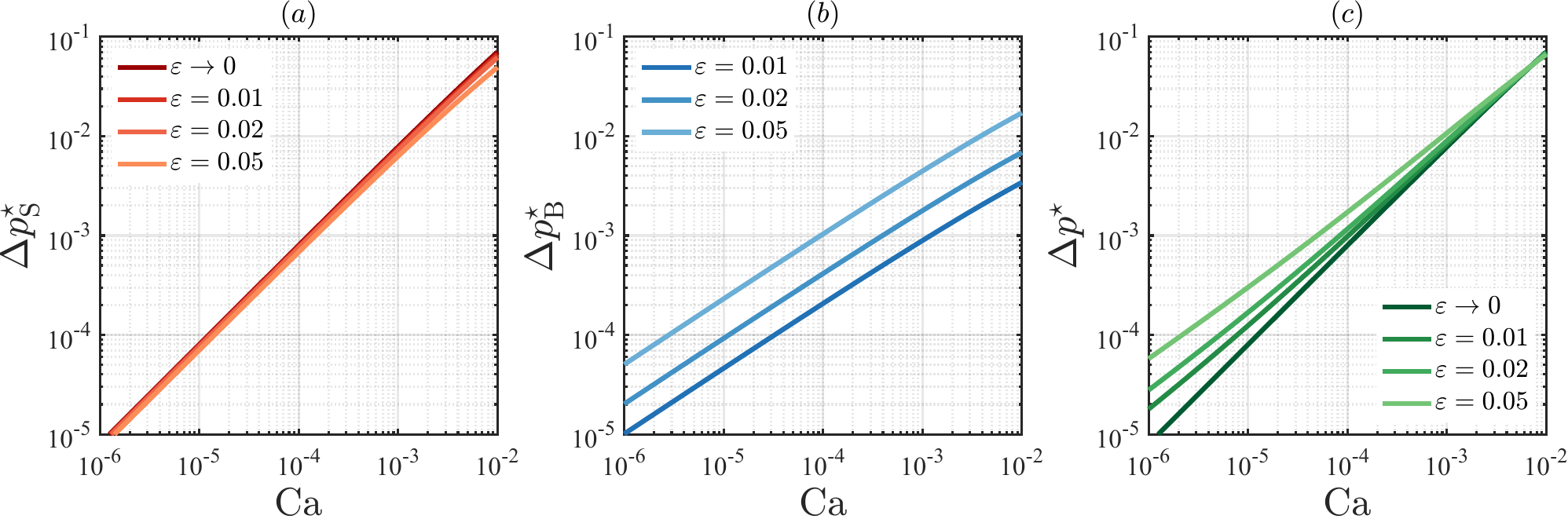}\\[1em]
    \includegraphics[width=1\linewidth]{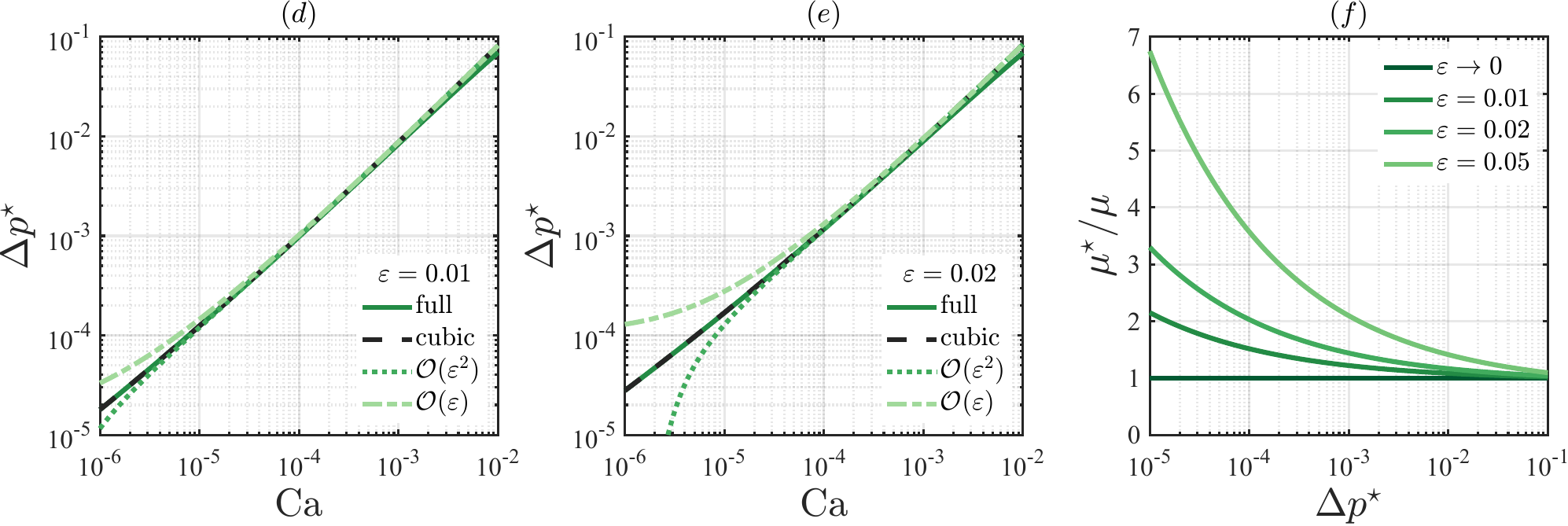}
    \caption{Dimensionless pressure drops as functions of the capillary number based on the bubble speed for different values of the perturbation parameter $\varepsilon$, see Eq.~\eqref{eq:new_cubicgamma}, for $\alpha=1$: ($a$) slug-related $\Delta{p}_{\rm{S}}^{\star}$, ($b$) bubble-related $\Delta{p}_{\rm{B}}^{\star}$, and ($c$) combined $\Delta{p}^{\star}$ contributions. 
    ($d$),\,($e$) Comparison of the solution to the full problem given in Eq.~\eqref{eq:new_cubicgamma} and to the cubic equation~\eqref{eq:cubic_a23}, with its asymptotic expansions, see Eq.~\eqref{eq:exp_lowCa}, including terms up to $\mathcal{O}(\varepsilon^2)$ and $\mathcal{O}(\varepsilon)$ for ($d$) $\varepsilon=0.01$ and ($e$) $\varepsilon=0.0{2}$. 
    ($f$) Effective viscosity of the two-phase flow in dimensionless terms, $\mu^{\star}/\mu$, see Eq.~\eqref{eq:mustar}, as a function of the dimensionless pressure drop $\Delta{p}^{\star}$ for different values of $\varepsilon$. 
    }
    \label{fig:asymptotics}
\end{figure}

To examine how a single elongated bubble causes deviations from Darcy-type behavior, we analyze the departure from the single-phase limit using an asymptotic approach. 
We consider a slender capillary tube, characterized by the small parameter
\begin{linenomath}
    \begin{equation}\label{eq:epsilon}
        \varepsilon=\dfrac{r}{L}\ll1,
    \end{equation}
\end{linenomath}
defined as the inverse of the slenderness ratio given in Eq.~\eqref{eq:newLast}, i.e., $\varepsilon=1/L^{\ast}$. 
We follow a dominant balance approach~\citep{Hinch1991, Bender1999} to determine the relative magnitude of each pressure contribution in Eq.~\eqref{eq:new_dim_p}. To this end, we introduce the capillary pressure scale as $\mathcal{P}_{{\alpha}}=\sigma/\left(\varepsilon^{{\alpha}}\,{r}\right)$, where the exponent $\alpha$ controls the reference length scale, thereby selecting the dominant physical mechanism.
The case $\alpha=0$ corresponds to the scaling considered in Sec.~\ref{sec:tube}.

In the following, we denote the dimensionless quantities obtained using the capillary pressure scale $\mathcal{P}_{\alpha}$ by the star, i.e., $\Delta{p}^{\star}=\Delta{p}\,\mathcal{P}^{-1}_{{\alpha}}$, and introduce the reduced capillary number $z$, defined in Eq.~\eqref{eq:zCa}. Considering a single bubble ($N=1$) of the minimum allowed length $b$, the normalization of Eq.~\eqref{eq:new_dim_p} yields
\begin{linenomath}
    \begin{equation}\label{eq:new_cubicgamma}
        {
        \Delta{p}^{\star}=\underbrace{\dfrac{8}{3}\,\varepsilon^{\alpha-1}\left(1-\varepsilon\,\dfrac{b}{r}\right)z^3\,\dfrac{v}{u}}_{\Delta{p}_{\rm{S}}^{\star}}+\underbrace{\varepsilon^{\alpha}\left(\beta_{2}\,z^2+\beta_{3}\,z^3\right)}_{\Delta{p}_{\rm{B}}^{\star}}.
        }
    \end{equation}
\end{linenomath}
To ensure that the bubble-related pressure drop, $\Delta{p}_{\rm{B}}^{\star}$, appears as $\mathcal{O}(\varepsilon)$ correction to the pressure drop in the liquid slugs, $\Delta{p}_{\rm{S}}^{\star}$, we choose $\alpha=1$ and we plot the pressure drop contributions in Figs.~\ref{fig:asymptotics}~($a$)\,--\ref{fig:asymptotics}\,($c$).
For small capillary numbers, since the velocity ratio $v/u$ approaches unity, see Eqs.~\eqref{vuw_relation} and~\eqref{hinfdivr}, and $b/r=b^{\ast}\to2$, see Eq.~\eqref{eq:bastmin}, the pressure drop model, Eq.~\eqref{eq:new_cubicgamma}, reduces to the following cubic equation in $z$:
\begin{linenomath}
    \begin{subequations}\label{eq:cubic_a23}
        \begin{align}
            \left(\frac{8}{3}+\varepsilon\,\hat{\beta}_{3}\right)& z^{3} + 
            \,\varepsilon\,\beta_{2}\,z^{2} - \Delta{p}^{\star}=0, \qquad \text{for}\quad z^2\ll1, \quad\text{where}\\
            & \hat{\beta}_{3}=\beta_{3}-\frac{16}{3}{\approx-9.724}.
        \end{align}
    \end{subequations}
\end{linenomath}
The coefficients $\beta_{2}$ and $\hat{\beta}_{3}$ capture the curvature- and normal stress-related contributions, respectively, to the bubble-induced pressure drop, see Eq.~\eqref{dpb_dim}.
In the limit of large scale separation ($\varepsilon\ll1$), the solution to the cubic equation~\eqref{eq:cubic_a23} -- derived via Cardano’s method but omitted here for brevity -- is well approximated using the first terms $\Lambda_{1},\,\Lambda_{2}$ of its power-series expansion:
\begin{linenomath}
    \begin{subequations}\label{eq:exp_lowCa}
        \begin{align}
            &\hspace{2.5cm}\text{Ca}\sim\dfrac{\Delta{p}^{\star}}{8}
            \left[1-\varepsilon\,\Lambda_{1}+\varepsilon^{2}\,\Lambda_{2}+\mathcal{O}(\varepsilon^{3})\right],
            \quad{\text{with}}
            \\
            \Lambda_{1}&=\dfrac{3\,{\hat{\beta}_{3}}}{8}+\dfrac{3^{2/3}\,{\beta_{2}}}{{4}\left(\Delta{p}^{\star}\right)^{1/3}},
            \qquad
            \Lambda_{2}=\dfrac{9\,{\hat{\beta}_{3}^{2}}}{64}+\dfrac{3^{1/3}\,{\beta_{2}^{2}}}{{8}\left(\Delta{p}^{\star}\right)^{2/3}}
            + 5\,\dfrac{3^{2/3}\,{\beta_{2}\,\hat{\beta}_{3}}}{{32}\left(\Delta{p}^{\star}\right)^{1/3}}{.}
        \end{align}
    \end{subequations}
\end{linenomath}
As the scale separation decreases, however, such approximation loses accuracy. In fact, for small values of the dimensionless pressure drop $\Delta{p}^{\star}$, higher-order terms grow rapidly, and additional terms are required for the series to converge to the full solution, see Fig.~\ref{fig:asymptotics}~($d$) and~\ref{fig:asymptotics}\,($e$).
Notably, even if the dimensionless pressure drop-flow rate relation, Eq.~\eqref{eq:exp_lowCa}, is nonlinear, at the leading order it can be written in a Darcy-like form by introducing a pressure-dependent effective viscosity $\mu^{\star}$, see Fig.~\ref{fig:asymptotics}\,($f$), and returning to dimensional variables:
\begin{linenomath}
    \begin{equation}\label{eq:mustar}
        Q\sim\dfrac{\pi\,r^4\,\Delta{p}}{8\,\mu^{\star}\,L}
        {,\quad\text{with}\quad}\mu^{\star}\!\left(\Delta{p}\right)= \mu\left[1+\dfrac{r}{L}\,\Lambda_{1}\!\left(\Delta{p}\right)\right].
    \end{equation}
\end{linenomath}

\section{\label{sec:bundle}Capillary bundle model}
In this section, we incorporate the single-capillary model of intermittent flow discussed in Secs.~\ref{sec:STM} and~\ref{sec:R&D} into a capillary bundle, i.e., an array of parallel, noninteracting capillary tubes of equal length subject to the same pressure drop.
This classical model~\citep{Scheidegger1954, Scheidegger1974, Vradis1993, Bartley1999, Bartley2001, Dahle2005, Yang2009, Civan2009, hansen2015fiber, Liu2016, Xiong2017, Zhao2020, Lanza2022} provides a simplified representation of flow in porous media, where structural heterogeneity is introduced through variations in pore-scale properties (e.g., capillary threshold pressures~\citep{Roy2019}, wetting angles~\citep{Fyhn2021}, pore sizes~\citep{Roy2021}) drawn from a given distribution. Our goal is to investigate how pore-size variability influences the macroscopic flow response of bubble trains in a capillary bundle and compare the resulting pressure–flow nonlinearity with the effective rheology of two-phase flow in porous media reported in the literature.

\begin{figure}
    \centering
    \includegraphics[width=0.8\linewidth]{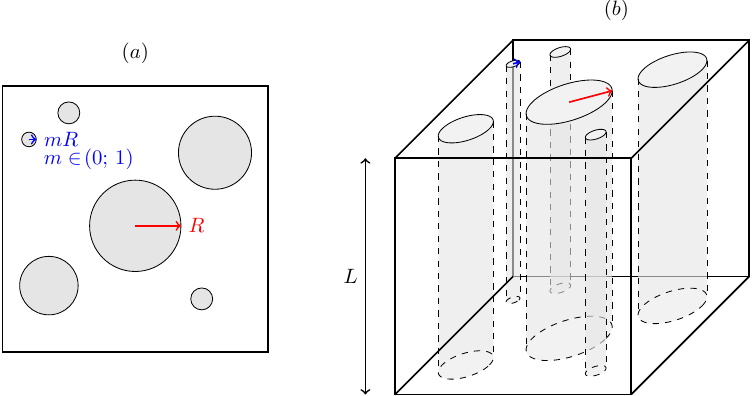}
    \caption{Schematic representation of a capillary bundle of length $L$, consisting of $M=6$ tubes with radii varying from a minimum value $m\,R$ to a maximum value $R$. ($a$) Top view. ($b$) 3D view.}
    \label{fig:CFBM}
\end{figure}

To do so, we consider a bundle of $M$ tubes whose radii are drawn from a probability density function $\rho(r)$ over the interval $[m R, R]$, as illustrated in Fig.~\ref{fig:CFBM}.
A finite lower cut-off $m$ ($0 < m < 1$) ensures well-defined flow by excluding vanishing radii.
Accordingly, we adopt a truncated power-law distribution with exponent $\gamma$:
\begin{linenomath}
    \begin{equation}\label{eq:distribRho}
        \rho(r;\,R,\,m,\,\gamma) =
                \begin{cases}
                    \dfrac{1-\gamma}{R^{1-\gamma}\left(1-m^{1-\gamma}\right)}\,r^{-\gamma}
                    & \text{if}\quad {\gamma\neq1,}\quad m R \le r \le R \\
                    \dfrac{1}{{\ln{(1/m)}}}\,r^{-1} 
                    & \text{if}\quad {\gamma=1,}\quad m R \le r \le R \\
                    0 &  {\text{if}\quad\, r < mR \quad\,\text{or}\quad\, r > R}.
                \end{cases}
    \end{equation}
\end{linenomath} 
As shown in the insets of Figs.~\ref{fig:QmstarDpstar_small} and~\ref{fig:QmstarDpstar_medium}, the exponent $\gamma$ controls the relative weight of small and large radii, favoring smaller pores for $\gamma>0$ and larger ones for $\gamma<0$. 
The distribution given by Eq.~\eqref{eq:distribRho} becomes uniform for $\gamma=0$, while the case $\gamma=1$ corresponds to the reciprocal (or log-uniform) distribution.

The single-capillary liquid flow rate for a given capillary tube of radius $r$ is
\begin{linenomath}
    \begin{equation}
        {
        q(r)=\pi\,r^2\,v(r),\quad\text{where}\quad v(r)=\dfrac{\sigma}{\mu}\,\text{Ca}_{l}(r).
        }
    \end{equation}
\end{linenomath}
This expression is based on the average speed of the liquid and implicitly accounts for the intermittent two-phase flow pattern. Hence, the steady-state liquid flow rate, averaged over the bundle, reads
\begin{linenomath}
    \begin{equation}\label{eq:Qm_dim}
        Q_{M}=\int_{m R}^{R}\,q(r)\,\rho(r)\,\mathrm{d}r.
    \end{equation}
\end{linenomath}

\subsection{\label{sec:bundle_adim}Dimensionless formulation and single-phase limit}
\begin{figure}
    \centering
    \includegraphics[width=1\linewidth]{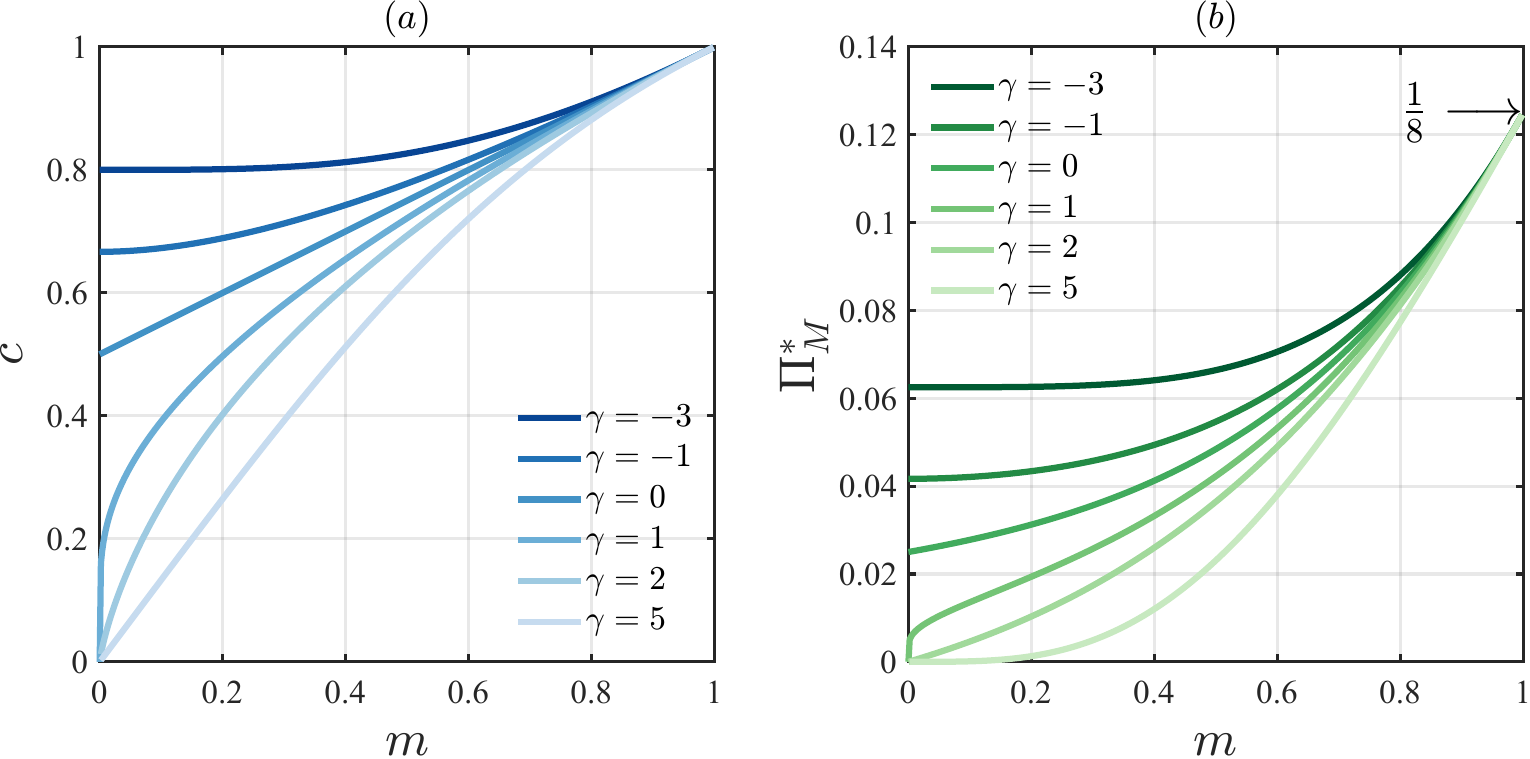}
    \caption{Variation of ($a$) the constant $c$ defined in Eq.~\eqref{eq:barr} and ($b$) the single-phase dimensionless permeability $\Pi_{M}^{*}$ defined in Eq.~\eqref{eq:RelPerm} as functions of the lower cutoff $m$, for different values of the exponent $\gamma$ of the power-law pore-size distribution given in Eq.~\eqref{eq:prob_dens_adim}.
    }
    \label{fig:Constantc_RelPerm}
\end{figure}

We develop a dimensionless formulation of the capillary bundle model by choosing the average radius of the distribution defined in Eq.~\eqref{eq:distribRho} as the reference length scale:
\begin{linenomath}
    \begin{equation}\label{eq:barr}
        \bar{r}=c\,R, \quad\text{where}\quad
        c\left(\gamma,\,m\right)=\begin{cases}
                        \dfrac{m-1}{\ln{m}} & \text{if}\quad\gamma=1\\
                        \dfrac{m\,\ln{m}}{m-1} & \text{if}\quad\gamma=2\\
                        \dfrac{1-\gamma}{2-\gamma}\,\dfrac{1-m^{2-\gamma}}{1-m^{1-\gamma}} & \text{otherwise}.
        \end{cases}
    \end{equation}
\end{linenomath}
Figure~\ref{fig:Constantc_RelPerm}\,($a$) shows the dependence of the constant $c$ in Eq.~\eqref{eq:barr} on the distribution parameters. All curves converge to $c = 1$ as $m \to 1$, as the radii distribution tends to a Dirac delta function (unit impulse) centered at $R$, consistent with the limit of homogeneous bundle. For $\gamma > 0$, $c$ decreases toward zero as $m \to 0$ due to the dominance of small tubes, while for $\gamma < 0$, $c$ exhibits a sublinear decrease with decreasing $m$, reflecting the stronger contribution of large tubes to the average radius.
Accordingly, the dimensionless radius and probability distribution read
\begin{linenomath}
    \begin{equation}\label{eq:prob_dens_adim}
        {r^{\ast}=\dfrac{r}{\bar{r}}\in\left[\dfrac{m}{c},\,\dfrac{1}{c}\right]}
        \quad\,\,
        {\text{and}}
        \quad\,\,
        \rho^{\ast}(r^{\ast})={\bar{r}\,\rho=}
        \begin{cases}
                        \dfrac{1}{{\ln{(1/m)}}}\,\left(r^{\ast}\right)^{-1} & \text{if}\quad\gamma=1\\
                        \dfrac{c^{1-\gamma}\left(1-\gamma\right)}{1-m^{1-\gamma}}\left(r^{\ast}\right)^{-\gamma} & \text{otherwise}.
        \end{cases}
    \end{equation}
\end{linenomath}

Then, we define the dimensionless pressure drop across the bundle and its slenderness ratio as
\begin{linenomath}
    \begin{equation}\label{eq:new_scale}
    {
        \Delta{p}^{\ast}=\dfrac{\Delta{p}}{\sigma/R},\qquad
        L^{\ast}=\dfrac{L}{R}.
        }
    \end{equation}
\end{linenomath}
The largest radius $R$ is used as the reference length scale in Eq.~\eqref{eq:new_scale} to obtain distribution-independent variables, enabling direct comparison between different bundles. Using Eqs.~\eqref{eq:prob_dens_adim} and~\eqref{eq:new_scale}, the single-capillary model of intermittent flow, Eq.~\eqref{eq:new_dim_p}, is written in dimensionless form as
\begin{linenomath}
    \begin{equation}
        {
        \Delta{p}^{\ast}
        =\underbrace{\dfrac{8}{3}\,z^3\,
            \dfrac{v}{u}\,L^{\ast}\left(1-\varphi\right)\left(\dfrac{1}{c\,r^{\ast}}\right)^{\!2}}
        _{\Delta{p}_{\rm{S}}^{\ast}}\,
        +
        \,\underbrace{\dfrac{N}{c\,r^{\ast}}
        \big(\beta_{2}\,z^2
        +\,\beta_{3}\,z^3\big)}_{\Delta{p}_{\rm{B}}^{\ast}},
        }
    \end{equation}
\end{linenomath}
where the reduced capillary number $z$ and the gas fraction $\varphi$ are defined as in~Sec.\,\ref{sec:tube}. Substituting the gas volume fraction defined in Eq.~\eqref{eq:phiPhi} and considering small capillary numbers gives the pressure drop model consistent at $\mathcal{O}(z^3)$:
\begin{linenomath}
    \begin{equation}\label{eq:dpast_solve}
        {
        \Delta{p}^{\ast}
        =\underbrace{\dfrac{8}{3}\,z^3\,
            L^{\ast}\left(1-\Phi-\dfrac{2}{3}\,N\,\dfrac{c\,r^{*}}{L^{\ast}}\right)\left(\dfrac{1}{c\,r^{\ast}}\right)^{\!2}}
        _{\Delta{p}_{\rm{S}}^{\ast}}\,
        +
        \,\underbrace{\dfrac{N}{c\,r^{\ast}}
        \big(\beta_{2}\,z^2
        +\,\beta_{3}\,z^3\big)}_{\Delta{p}_{\rm{B}}^{\ast}},
        \quad \text{for}\quad z^2\ll1.}
    \end{equation}
\end{linenomath}
Finally, by using Eq.~\eqref{eq:Qm_dim}, we express the dimensionless liquid flow rate as
\begin{linenomath}
    \begin{equation}\label{eq:QMQMast}
        Q_{M}^{\ast}={\dfrac{Q_{M}}{\pi\,R^{2}\,\sigma/\mu}=\,}
        \displaystyle{c^{2}\int_{m/c}^{1/c}\left(r^{\ast}\right)^{2}\,\text{Ca}_{l}(r^{\ast})\,\rho^{\ast}(r^{\ast})\,\mathrm{d}r^{\ast}}.
    \end{equation}
\end{linenomath}
The flow rate corresponding to a given pressure drop is obtained by combining Eqs.~\eqref{eq:dpast_solve} and~\eqref{eq:QMQMast}, noting that $\text{Ca}_{l}\sim\text{Ca}=z^3/3$ for $z^2\ll1$, see Eq.~\eqref{eq:CalCa}.
Alternatively, the capillary number could be defined using the Darcy velocity -- i.e., the volumetric flow rate per total cross-sectional area~\citep{Nield2006book}-- although this choice requires specifying the bundle's porosity as an additional parameter.
In the absence of bubbles, i.e., $\Phi=0$ and $N=0$, the single-phase flow condition is recovered, yielding the following closed-form expression for the dimensionless permeability of the bundle: 
\begin{linenomath}
    \begin{equation}\label{eq:RelPerm}
        \Pi_{M}^{*}=\dfrac{Q_{M}^{\ast}}{\Delta{p}^{\ast}/{L^{\ast}}}=
        \begin{cases}
            \dfrac{m^4-1}{32\,\ln{m}}
            & \quad\text{if}\quad\gamma=1\\
             \dfrac{1}{2}\,\dfrac{m^4\,\ln{m}}{m^4-1}       & \quad\text{if}\quad\gamma=5\\
            \dfrac{1}{8}\,\dfrac{1-\gamma}{5-\gamma}\,\dfrac{1-m^{5-\gamma}}{1-m^{1-\gamma}}
            & \quad\text{otherwise}.
        \end{cases}
    \end{equation}
\end{linenomath}
As shown in Fig.~\ref{fig:Constantc_RelPerm}\,($b$), in the limit of homogeneous bundle ($m\to1$), the dimensionless permeability converges to $1/8$, consistent with Darcy flow through a single round capillary.
Any heterogeneous power-law distribution of pore sizes yields a lower permeability, with $\Pi_{M}^{*}$ increasing monotonically with $m$ for any $\gamma$. 
As $m\to0$, two distinct behaviors emerge due to the inclusion of infinitesimally small pores: for $\gamma\ge1$, the permeability vanishes, reflecting flow blockage, while for $\gamma<1$ the bundle retains a finite permeability, i.e., $\Pi_{M}^{*}\to\dfrac{1}{8}\,\dfrac{1 - \gamma}{5 - \gamma}$.

In the two-phase regime, the numerical evaluation of the pressure drop-flow rate relation (see Sec.~\ref{sec:2phasebundle}) enables us to establish an analogy between the nonlinear response predicted by the simplified capillary bundle model of intermittent flow and the effective rheology of two-phase flow in porous media, typically expressed as a power-law, $\Delta p \propto Q_{M}^{\zeta}$, as given by Eq.~\eqref{eq:NonDarcyNew} (with $\Delta{p}_{t}=0$).
Next, we compute the exponent $\zeta$ predicted by our simplified capillary bundle model, analyze its dependence on the flow parameters, and compare it with values reported in the literature, see Table~\ref{tab:literature}.

\subsection{\label{sec:2phasebundle}Analysis of the bundle two-phase pressure-flow response}
We characterize the bulk behavior of the two-phase bubble-train flow through the capillary bundle by examining the influence of the pore-size distribution and phase topology on the flow rate-pressure curve $Q_{M}^{\ast}\left(\Delta{p}^{\ast}\right)$. This procedure can be summarized as follows:
\begin{enumerate}
    \item \label{item_step_ISP} 
    Initializing the statistical properties of the bundle (in terms of the distribution parameters $m$ and $\gamma$) and computing the constant $c$ according to Eq.~\eqref{eq:barr}.
    
    \item \label{item_step_DCC} 
    Selecting the dimensionless length of the bundle $L^{{\ast}}$ and the gas-phase topology (in terms of saturation  $\Phi$ and number of bubbles $N$) at a fixed pressure drop $\Delta{p}^{\ast}$.
    
    \item \label{item_step_CNM} 
    Mapping the reduced capillary number $z$ over the range of dimensionless radii defined by $m/c\le{r^{\ast}}\le1/c$ by solving Eq.~\eqref{eq:dpast_solve} iteratively.
    
    \item \label{item_step_CNC}
    Computing the corresponding function $\text{Ca}_{l}\left(r^{\ast}\right){\sim{z^3}/3}$, consistent with Eq.~\eqref{eq:CalCa}.
    
    \item \label{item_step_FRC}
    Computing the corresponding flow rate $Q_{M}^{\ast}$ through numerical evaluation of the integral in Eq.~\eqref{eq:QMQMast}.
    
    \item \label{item_step_IPD}
    Repeating steps \eqref{item_step_CNM}\,-\,\eqref{item_step_FRC} for increasing pressure drops.
\end{enumerate}
To characterize the resulting pressure drop-flow rate nonlinearity, the logarithmic derivative
\begin{linenomath}
    \begin{equation}\label{eq:derivLog}
        \zeta\left({Q_{M}^{\ast}}\right)=\dfrac{\mathrm{d}\log_{10}{\Delta{p}^{\ast}}}{\mathrm{d}\log_{10}{Q_{M}^{\ast}}}
    \end{equation}
\end{linenomath}
is computed pointwise from the data using a finite-difference approximation.

\begin{figure}
        \centering
    \includegraphics[width=\linewidth]{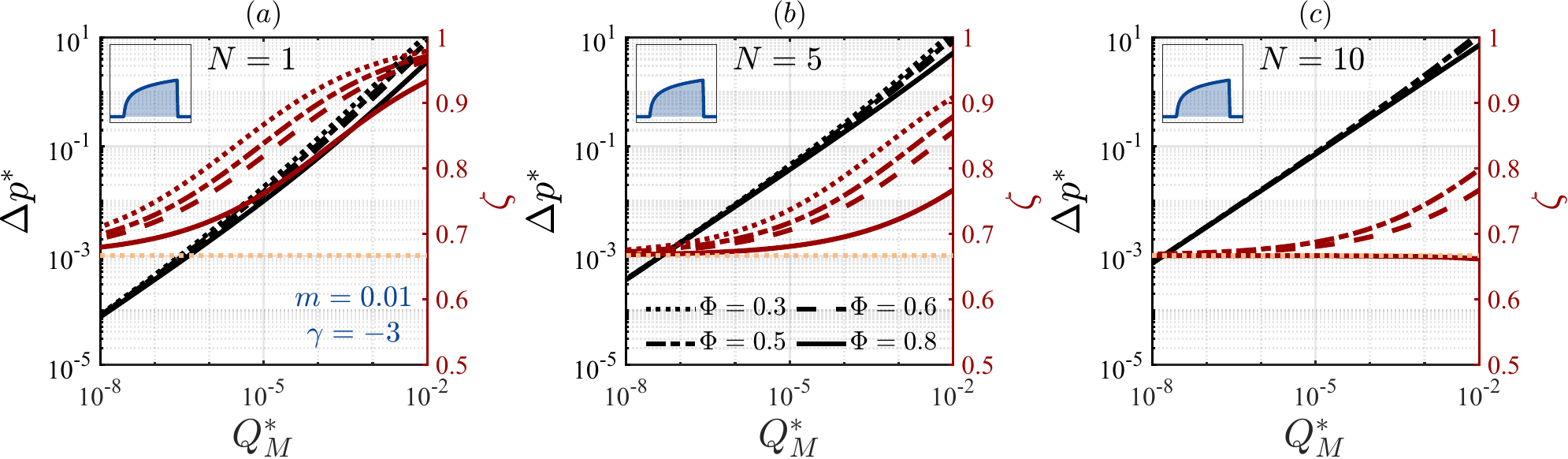}\\[1em]
    \includegraphics[width=\linewidth]{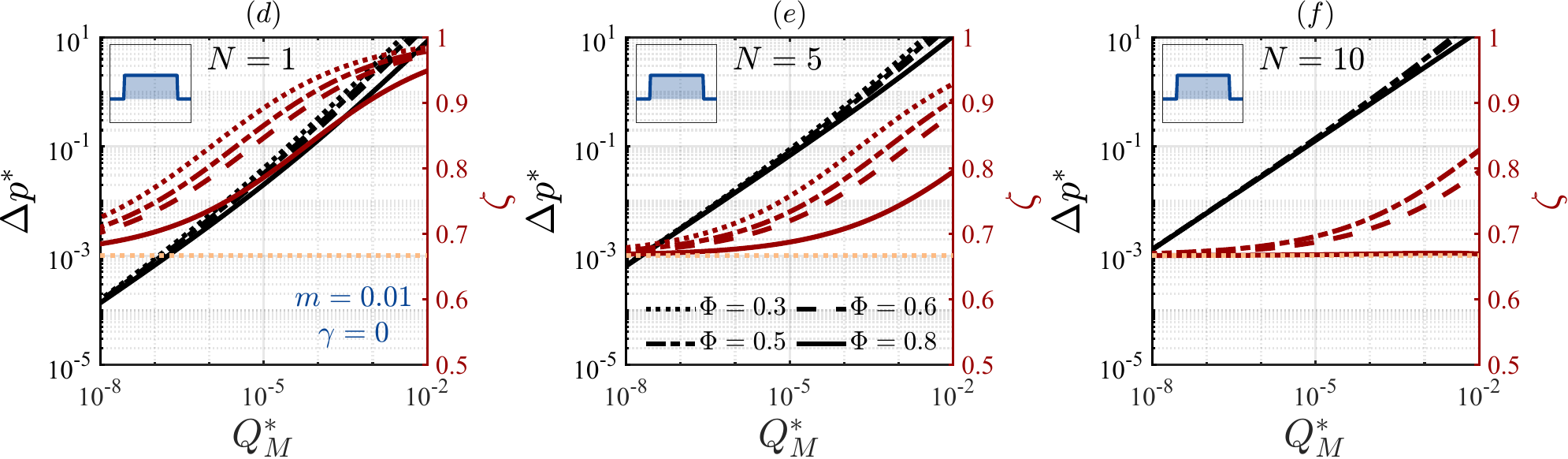}\\[1em]
    \includegraphics[width=\linewidth]{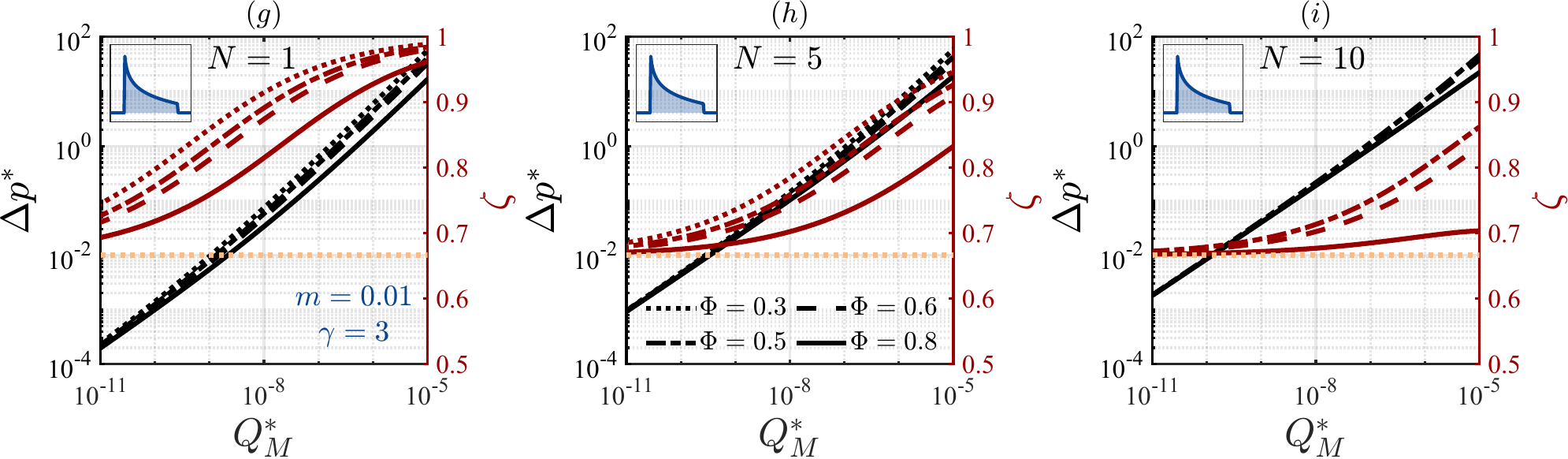}
    \caption{Evolution of the dimensionless pressure drop, $\Delta p^{\ast}$ (left ordinate), and the flow exponent, $\zeta$ (right ordinate), as functions of the dimensionless flow rate, $Q_{M}^{\ast}$. Rows correspond to pore-radii distribution exponents $\gamma$, columns to the number of bubbles $N$, and linestyles indicate the gas saturation $\Phi$. Each subplot includes an inset illustrating the pore-size distribution. Results corresponds to a capillary bundle with slenderness ratio $L^{\ast} = 100$ and lower cutoff $m = 0.01$.}
    \label{fig:QmstarDpstar_small}
\end{figure}

\begin{figure}
        \centering
    \includegraphics[width=\linewidth]{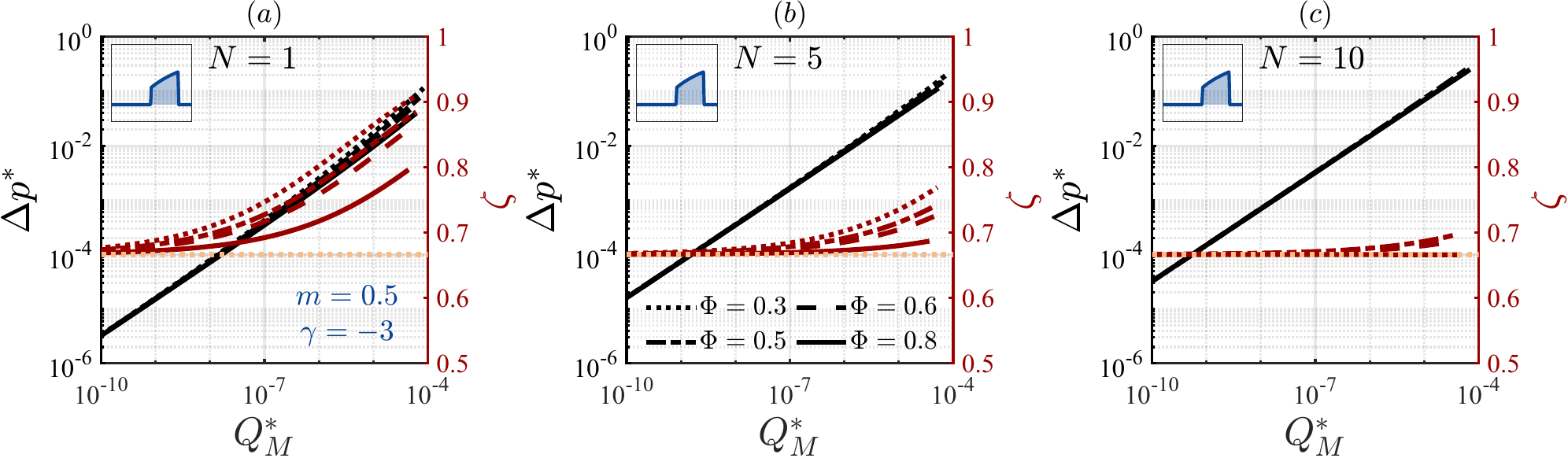}\\[1em]
    \includegraphics[width=\linewidth]{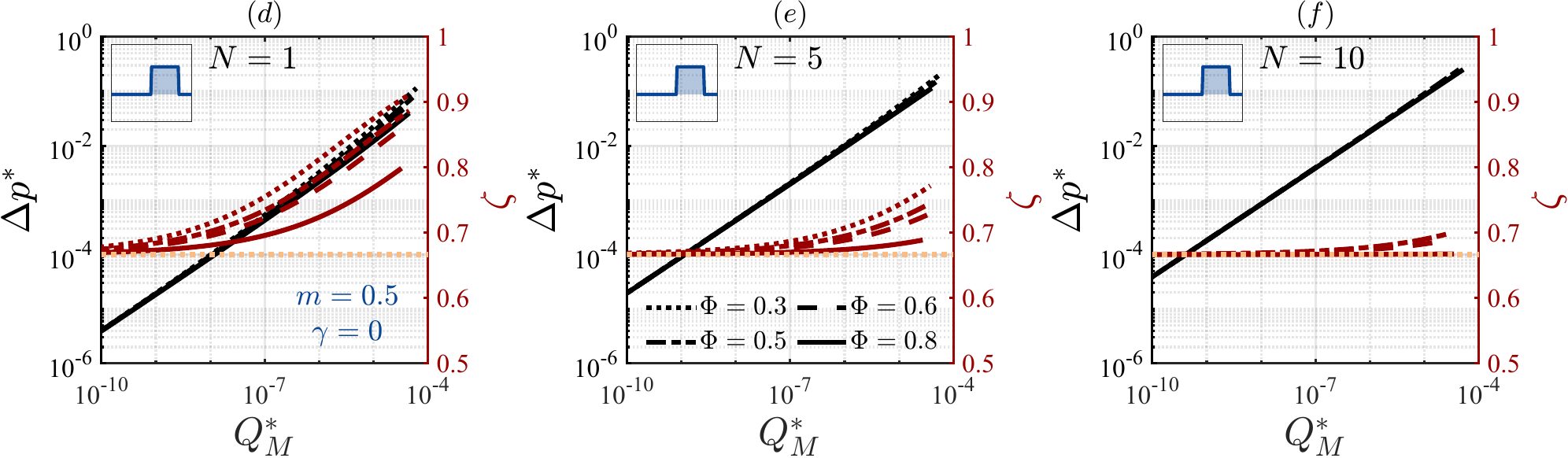}\\[1em]
    \includegraphics[width=\linewidth]{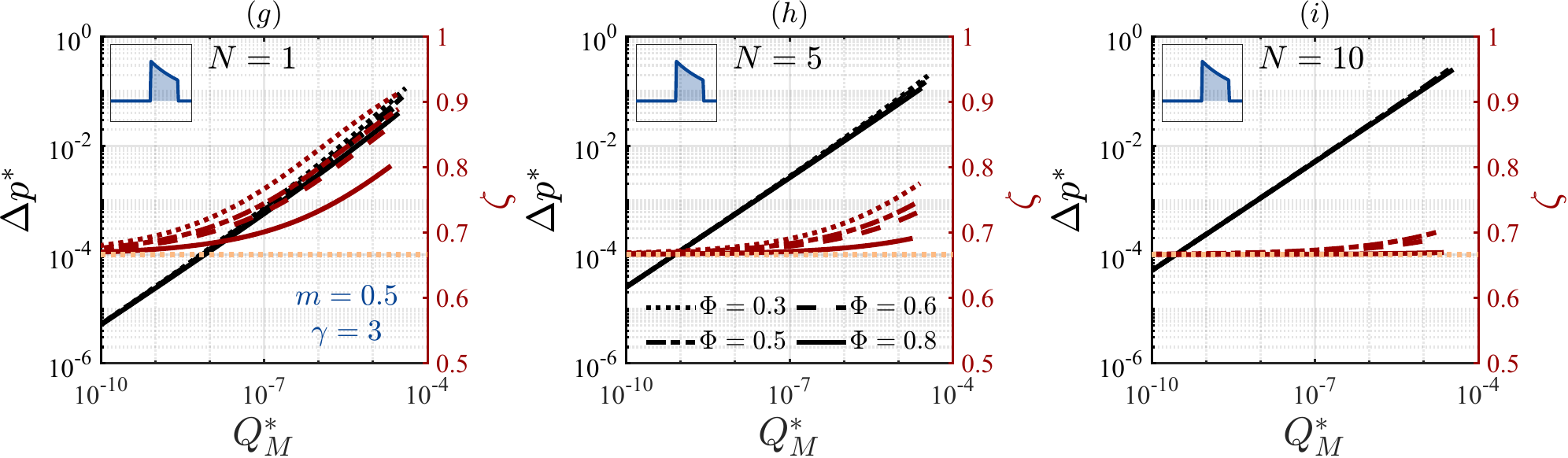}
    \caption{Evolution of the dimensionless pressure drop, $\Delta p^{\ast}$ (left ordinate), and the flow exponent, $\zeta$ (right ordinate), as functions of the dimensionless flow rate, $Q_{M}^{\ast}$. Rows correspond to pore-radii distribution exponents $\gamma$, columns to the number of bubbles $N$, and linestyles indicate the gas saturation $\Phi$. Each subplot includes an inset illustrating the pore-size distribution. Results corresponds to a capillary bundle with slenderness ratio $L^{\ast} = 100$ and lower cutoff $m = 0.5$.}
    \label{fig:QmstarDpstar_medium}
\end{figure}

We analyze the effective rheology of the two-phase flow through the capillary bundle by assessing how different pore-size distributions affect the nonlinearity of the pressure drop-flow rate relation.
The results shown in Figs.~\ref{fig:QmstarDpstar_small} and~\ref{fig:QmstarDpstar_medium} correspond to lower cutoffs of the distribution equal to $m=0.01$, and $m=0.5$, respectively, while the slenderness ratio of the bundle is set to $L^{\ast}=100$. This choice enables us to model structural heterogeneity through the inclusion or exclusion of smaller pores.  
Each figure comprises a $3\times 3$ panel grid, where rows correspond to different exponents of the truncated power-law distribution, i.e., $\gamma=-3$, $0$ and $3$, as indicated in the insets, whereas columns represent increasing number of bubbles per tube, i.e., $N=1$, $N=5$, and $N=10$.
Different linestyles within each panel denote the bubble-train volume fraction, selected within the applicability range of the single-capillary model discussed in~Sec.\,\ref{sec:applicability}. 
The results are presented in dimensionless form as pressure drop-flow rate curves $\Delta p^{\ast}(Q_{M}^{\ast})$ (left ordinate), together with the local scaling exponent $\zeta(Q_{M}^{\ast})$, defined in Eq.~\eqref{eq:derivLog} (right ordinate); the benchmark value predicted by~\citeauthor{Bretherton1961} {for a single elongated bubble} (i.e., $\zeta = 2/3$) is indicated by a light red dashed line.

When smaller radii are included in the bundle (Fig.~\ref{fig:QmstarDpstar_small}), the exponent of the pore-size distribution strongly controls the order of magnitude of the resulting flow rate. In particular, when the distribution favors smaller radii (third row), the flow rate is up to three orders of magnitude lower than in the case of uniform distributions (second row) or distributions weighted toward larger radii (first row). For a single bubble ($N=1$), the local logarithmic slope exhibits a smooth transition from the Bretherton regime (i.e., $\zeta=2/3$) to a Darcy-type response (i.e., $\zeta=1$) as the pressure drop increases.
Increasing both the number of bubbles and the gas volume fraction progressively flattens the curves towards $\zeta=2/3$. This trend reflects the enhanced influence of the dispersed phase, as the progressive replacement of liquid slugs by bubbles mitigates the viscous contribution of the carrier phase and dampens the overall rheological response. 
For narrower distributions centered around intermediate radii (Fig.~\ref{fig:QmstarDpstar_medium}), the rheological response becomes less sensitive to the distribution exponent. The curves are more tightly clustered, indicating reduced variability in flow behavior due to the exclusion of smaller pores. A similar trend is observed for even narrower distributions centered around larger radii, corresponding to more homogeneous bundles (not shown for brevity).

Interestingly, the flow exponent predicted by our simplified capillary bundle model of bubble-train flow is qualitatively similar to those reported in the literature for two-phase flow in porous media (see Tab.~\ref{tab:literature}). Specifically, our findings closely align with those of~\citet{Roy2019}, particularly in regimes where the pressure drop is dominated by the dispersed phase (e.g., high gas saturation or number of bubbles), where our model recovers a flow exponent $\zeta=2/3$. Our analysis complements classical capillary-bundle models by accounting for the film-related hydrodynamics.
However, direct comparison between our simplified bubble-train system and two-phase flow in porous media should be made with caution. In fact, intermittent flow in real porous media involves substantial additional complexities, both geometric (pore interconnectivity and irregularity) and dynamic (pinning/de-pinning, splitting/merging of bubbles), which are not captured by our capillary-bundle model. A more detailed discussion of these limitations and their implications for interpreting our results is provided in the concluding section.

\section{\label{sec:conclusion}Conclusions}
In this work, we explored the nontrivial rheological response of two-phase intermittent flow in a single capillary tube and in an array of parallel, noninteracting tubes of different radii, i.e., a capillary bundle, focusing on the steady-state pressure drop-flow rate relation. 
Our approach relies on the decomposition of the total pressure drop into contributions from liquid slugs and bubble domains. 
In the liquid regions, the pressure drop scales linearly with the flow rate due to viscous effects, whereas in the bubble domains it follows a sublinear scaling with exponent $2/3$, consistent with~\citet{Bretherton1961} theory, reflecting the dominant capillary effects associated with thin-film dynamics.
This framework allowed us to characterize the emergence of nonlinear pressure drop-flow rate behavior in a capillary bundle as a function of gas saturation, number of bubble, and structural heterogeneity of the medium, which we parametrized though two independent descriptors of the power-law pore-size distribution.

Our simulations indicate that the largest pores primarily control the overall magnitude of the flow rate. Pore-size distributions that are moderately or strongly centered around the largest radii exhibit less sensitivity to asymmetries in the distribution tails, whereas distributions including the smallest pores lead to markedly different macroscopic responses.
Increasing the gas-phase contribution (both in terms of volume fraction and number of bubbles) dampens the effective rheological response of the system, promoting flow regimes characterized by a sublinear pressure drop-flow rate relation similar to that identified by~\citet{Bretherton1961} for a single elongated bubble. 

A key advance of our study is the incorporation of the film hydrodynamics typical of intermittent gas–liquid flow into a capillary bundle model. This enables us to interpret the effective two-phase behavior as an outcome of the interplay between pore-scale heterogeneity and the spatial configuration of the dispersed phase.
The flow exponents obtained are qualitatively consistent with those reported for two-phase flow in porous media. However, our model should be regarded as a minimal proxy for a porous structure. In fact, real porous media exhibit nonlinear behavior arising from multiple coupled mechanisms, including complex conductive pathways~\citep{Roux1987} that are neither straight nor uniform in cross-section and are typically interconnected.
To retain analytical tractability, we introduced several simplifying assumptions: (i) the fore and aft liquid regions adjacent to each bubble were treated as hydraulically independent so pressure drops are additive and (ii) the volume fraction and the number of bubbles were assumed uniform across all capillaries in the bundle; moreover, (iii) our capillary bundle model cannot capture pinning and depinning events, nor mixing processes (such as bubble splitting and merging), which are intrinsic to realistic porous networks.

Future work should therefore adapt the present pressure-drop formulation to more realistic network models that incorporate flow partitioning at pore junctions, such as interacting capillary bundles~\citep{Dong1998}.
Such approaches are essential to elucidate the origin of macroscopic flow nonlinearity in porous media, where phase redistribution and pathway competition strongly influence the pore-scale dynamics and, consequently, the local effective rheology~\citep{Ashraf2023, Deng2024}.

\begin{acknowledgments}
This work was partly supported by the Research Council of Norway through its Centers of Excellence funding scheme, Project No. 262644, and the INTPART program, Project No. 309139. 
A.H. and S.S. acknowledge funding from the European Research Council (Grant Agreement No. 101141323 AGIPORE).
D.P. acknowledges funding from the European Union `NextGenerationEU', Ministero dell'Università e della Ricerca (MUR) `Italiadomani' Piano Nazionale di Ripresa e Resilienza (PNRR), Mission 4, Research Project PRIN 2022 ``Predictive forecasting and risk assessment for CO\textsubscript{2} transport in pipelines'', MUR Code No.: 20229JPN53; CUP Master Code No.: J53D23002000006; and CUP Code No.: D53D23003250006. 
\end{acknowledgments}



\bibliography{apssamp}

\end{document}